\documentclass{aa}  

\usepackage{graphicx}
\usepackage{xcolor}
\usepackage{placeins}

\newcommand{\target}{{PKS 0023$-$26}}

\newcommand{\kms}{$\,$km$\,$s$^{-1}$}

\newcommand{\ergs}{$\,$erg$\,$s$^{-1}$}

\newcommand{\mJybeam}{mJy beam$^{-1}$}

\newcommand{\msun}{{$M_\odot$}}
\newcommand{\msunyr}{{$M_\odot$ yr$^{-1}$}}

\newcommand{\pks}{{PKS 0023$-$26}}
\newcommand{\coOne}{{CO(1-0)}}
\newcommand{\coTwo}{{CO(2-1)}}
\newcommand{\coThree}{{CO(3-2)}}

\newcommand{\rdt}{$R_{\rm 32}$}
\newcommand{\rto}{$R_{\rm 21}$}
\newcommand{\Tex}{$T_{\rm ex}$}

\def\emph#1{{\sl #1}}
\newcommand{\ltsima} {$\; \buildrel < \over \sim \;$}
\newcommand{\gtsima} {$\; \buildrel > \over \sim \;$}
\newcommand{\lta} {\lower.5ex\hbox{\ltsima}}
\newcommand{\gta} {\lower.5ex\hbox{\gtsima}}

\newcommand\chandra{Chandra}

\usepackage{txfonts}

\begin{document} 

   \title{The changing impact of radio jets as they evolve:\\The view from the cold gas}

\authorrunning{Oosterloo et al.}

\author{Tom Oosterloo\inst{1,2}, Raffaella Morganti\inst{1,2}, Clive Tadhunter\inst{3}, Aneta Siemiginowska\inst{4}, \\ Ewan O'Sullivan\inst{4}, Giuseppina Fabbiano\inst{4}}

\institute{ASTRON, the Netherlands Institute for Radio Astronomy, Oude Hoogeveensedijk 4, 7991 PD, Dwingeloo, The Netherlands. 
\and
Kapteyn Astronomical Institute, University of Groningen, Postbus 800,
9700 AV Groningen, The Netherlands
\and
Department of Physics and Astronomy, University of Sheffield, Sheffield, S7 3RH, UK
\and
Center for Astrophysics, Harvard \& Smithsonian, 60 Garden Street, Cambridge MA, 02138 USA
}
 \abstract
{We present ALMA CO(1-0) and CO(3-2) observations of a powerful young radio galaxy, \target, that is hosted by a far-infrared bright galaxy. The galaxy has a luminous optical active galactic nucleus (AGN) and a very extended distribution of molecular gas.  We used these observations (together with available ALMA CO(2-1) data) to trace the impact of the active nucleus across the extent of the radio emission and beyond on scales of a few kiloparsec (kpc). 
Despite the strength of the optical AGN, the kinematics of the cold molecular gas is strongly affected only in the central kpc, and it is more weakly affected around the northern lobe. We found other signatures of the substantial impact of the radio AGN, however.  Most notably, the extreme line ratios of the CO transitions in a region that is aligned with the radio axis indicate conditions that are very different from those observed in the undisturbed gas at large radii. The non-detection of \coOne\ at the location of the core of the radio source implies extreme conditions at this location.
Furthermore, on the scale of a few kpc,  the cold molecular gas  appears to be wrapped around the northern radio lobe. This suggests that a strong jet-cloud interaction has depleted the northern lobe of molecular gas, perhaps as a result of the hot wind behind the jet-induced shock that shreds the clouds via hydrodynamic instabilities. The higher gas velocity dispersion and molecular excitation that we observed close to this location may then be the result of a milder interaction in which the expanding jet cocoon induces turbulence in the surrounding interstellar medium.
These results highlight  that the impact of an AGN can manifest itself not only in the kinematics of the gas, but also in  molecular line ratios and in the distribution of the gas. The work also highlights that it is important to spatially resolve the gas throughout the radio source to trace different modes of AGN feedback that can coexist. 
Although the radio plasma and the cold molecular gas are clearly coupled, the kinetic energy that is transferred to the interstellar medium is only a small fraction of the energy available from the AGN.
 }
   \maketitle  


\section{Introduction}
\label{sec:introduction}

As a result of the wide range of sizes they cover, radio jets have the potential to impact the surrounding medium on scales from parsec (pc) to hundreds of kiloparsec (kpc),  from the  circumnuclear- and interstellar medium (ISM) to the intergalactic medium (IGM). On galaxy cluster scales, this has been shown by  studies of the cavities in the hot X-ray emitting gas that is created by the radio jets in cool-core clusters (bright cluster galaxies; BCG) and by measuring their lifting power (e.g. \citealt{McNamara16,Russell19,Tamhane22}).
On the other hand, we also know that radio jets can impact their surrounding medium already {   at an earlier stage} on small scales (pc to kpc) in the circumnuclear regions of galaxies.  In particular, young jets can drive fast outflows and increase the turbulence of the surrounding gas. This has been shown by several studies that traced warm ionised gas (e.g. \citealt{Holt08,Santoro20}) as well as the cold and warm molecular component (e.g. \citealt{Guillard12,Tadhunter14,Murthy22,Ruffa24,Cresci23,Costa24}).
As in the case of radiation-driven outflows, however, most of the most massive cold-gas outflows that are driven by radio jets have been found in regions that are limited to the inner kpc, and the connection of the impact of jets on small scales to what is seen on large scales is still unclear. 

The predictions from numerical simulations are interesting in this context. According to these simulations of jets that expand in a clumpy medium (\citealt{Sutherland07,Mukherjee16,Mukherjee18a,Tanner22,Dutta24}), 
the initial direct interaction of the jet with the cloud does not only produce   gas outflows, but also a cocoon of shocked material that is filled with expanding dense gas {around the jet}. Crucially, this cocoon also expands in a direction perpendicular to the jet, which ensures that a large volume of the ISM and not just the immediate surroundings of the jet is indirectly affected by the radio plasma. The energy is injected via the continual dissipation from the several interactions of the jets with the surrounding clouds, which drives the expansion of the cocoon and results in a high turbulence of the gas. 
The impact of this bubble may become  dominant when the radio source grows and breaks out from the central regions \citep{Mukherjee16,Mukherjee18a}. Although the characteristics of the bubble depend, among other factors, on the jet power (low-power jets interact for a longer time with the surrounding medium), the bubble of shocked gas can ensure that the energy released from the radio plasma is coupled to the ISM in low- and high-power jets. 
Observations have suggested that an expanding bubble of gas like this exists for a growing number of cases and for various phases of the gas \citep{Villar99,Croston07,Mingo11,Mingo12,Audibert19,Zovaro19,Girdhar24,Murthy25}, but the development of the bubble with the evolution of the jet remains to be clarified.

   \begin{figure}
   \centering
      \includegraphics[angle=0,width=8cm]{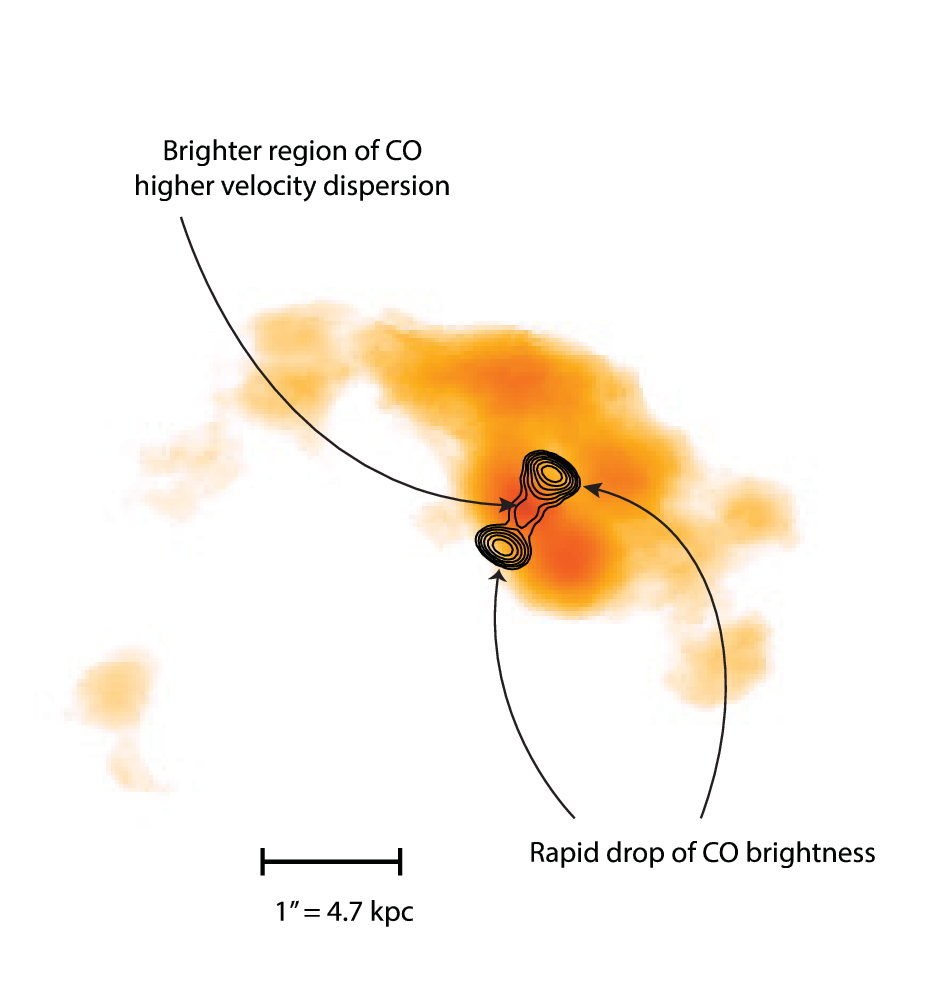}
   \caption{Distribution of the cold molecular gas (\coTwo) and of the radio continuum, modified from \citet{Morganti21}.  The contours represent the 1.7 mm (176~GHz)  continuum emission, derived from the uniformly weighted data. The contours are at 1.3, 2.6, 5.2, 10.4, 20.8, and 41.6 \mJybeam.}
              \label{fig:paper1}
    \end{figure}

A way to trace this interaction is to complement measurements of the gas kinematics  with those of its physical properties. 
The cold molecular gas (traced by CO) is ideally suited for studying these effects   because its distribution and kinematics can be observed at a high spatial resolution (see reviews by  \citealt{Bolatto13,Carilli13,Ruffa24}).   In addition, observations of multiple CO transitions and other molecules can inform us about the physical conditions of the molecular gas. For example, in regions in which the  jets/lobes (or an expanding jet cocoon) and the ISM interact, extreme line ratios have been observed. Examples include  IC~5063, PKS~1549--79, NGC~3100,  the Tea Cup, and B2~0258+35 (\citealt{Oosterloo17,Oosterloo19,Ruffa22,Audibert23,Murthy25}, respectively), and other cases in \cite{Audibert25}. 

In  the above cases,  the effects of the AGN on the gas are observed to extend to scales of a kpc or smaller. 
The object studied in this paper, \target, is particularly suited for extending these  investigations to  scales of several kpc because it is a young but relatively more evolved  radio galaxy with a size of a few kpc in which the radio jet might be in the process of breaking out from the gas-rich central region.
We therefore earlier observed \target\ with ALMA to explore the impact of young radio jets on kpc scales \citep{Morganti21}. These ALMA \coTwo\ observations  revealed a large amount of molecular gas  that was  estimated to be between $3.7 \times 10^9$ \msun\ and $3.1 \times 10^{10}$  \msun, depending on the assumed CO(2-1)/CO(1-0) line ratio and the CO-to-H$_2$ conversion factor (see  Fig.\ \ref{fig:paper1}). The observations  suggested an evolutionary scenario in which only the gas in the very inner region might be part of an outflow resulting from the direct interaction of the radio plasma jet and the ISM, while a gentler interaction occurs on larger scales (see Fig.\ 11 in \citealt{Morganti21}). 
These two different effects suggest that the impact of the jet might evolve as the jet expands. Interestingly, this has also been suggested to occur in  other types of AGN (i.e.\ radio-quiet AGN; e.g. \citealt{Girdhar24}).

   \begin{figure}
   \centering
\includegraphics[angle=0,width=9cm]{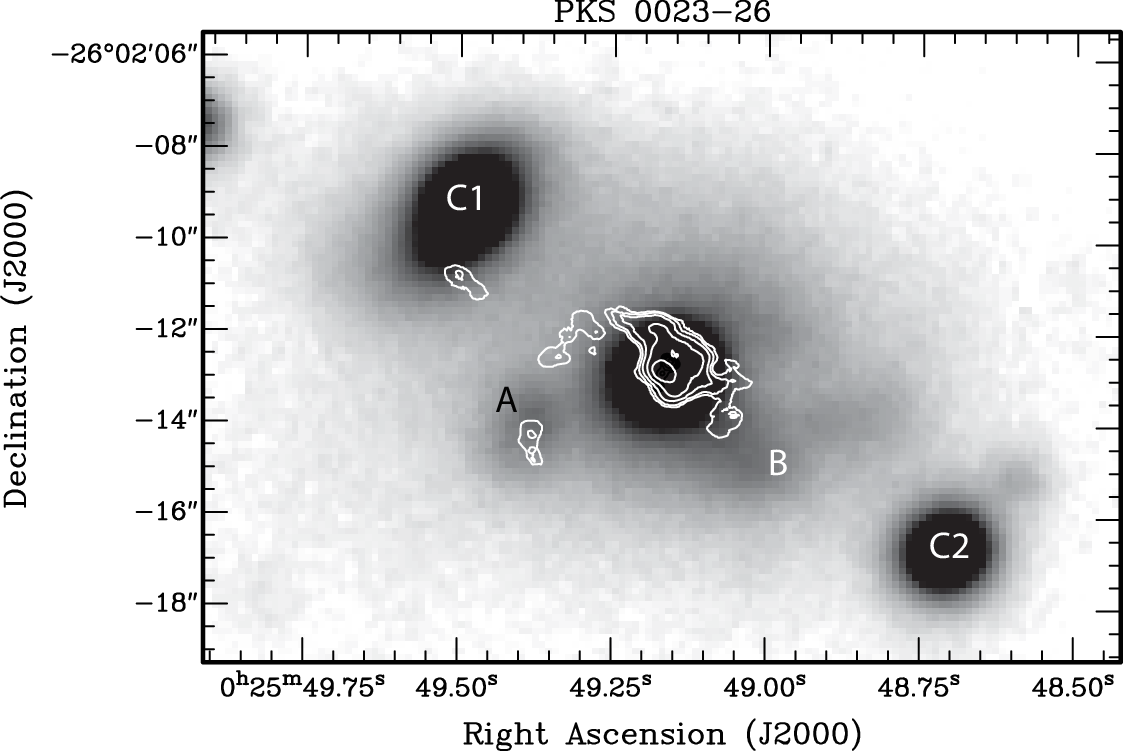}
   \caption{
   Optical image from Gemini \citep{Ramos11}. The contours of the \coThree\ emission are overplotted. Tails are seen toward nearby galaxies (projected distances $\sim$10 kpc). The galaxies labelled C1, C2, and C3 (outside the figure) are confirmed to have redshifts similar to that of \target\ \citep{Tadhunter11}. 
   }
              \label{fig:environment}
    \end{figure}

In order to test this scenario for \target\ and to understand how it may affect the conditions of the ISM, observations of multiple transitions of CO are needed. 
Furthermore, the origin of the high mass of cold molecular gas that was found with ALMA \citep{Morganti21} is also unclear, and  it is uncertain whether it might be connected with the rich, cluster-like environment. 
Therefore, we performed new ALMA observations that targeted two other CO transitions in order to trace the line ratios. We also performed new \chandra\ observations to investigate the environment in more detail. The Chandra observations will be presented in a companion paper  \citep{Siemiginowska2025}.

\begin{table*}
\caption{Parameters of the ALMA data cubes and millimetre images.}
\begin{center}
\begin{tabular}{lcr@{ $\times$ }lcrr@{ $\times$ }lccccc} 
\hline\hline 
   & Frequency & \multicolumn{3}{c}{Beam \& PA} & &\multicolumn{2}{c}{Beam}& Velocity resolution  & Noise     & Description\\
   &  (GHz)   & \multicolumn{2}{c}{(arcsec)}  & (deg)& &\multicolumn{2}{c}{(kpc)} & (\kms)          & (\mJybeam) & \\
\hline
\coOne\   &        & 0.39 &   0.26  & \phantom{+}84.8 &&1.8 & 1.2 & 33.6  &  0.08   & \\
\coTwo\   &        &  0.45&0.34 & --77.8 & & 2.1 & 1.6 &33.6    &  0.12   &  \\
\coThree\ &        & 0.29 &   0.25  & \phantom{+}78.3  & & 1.4 & 1.2 & 33.6  &  0.16   &   \\
\coOne\   &        &   0.19&  0.13& \phantom{+}57.2 & & 0.9 & 0.6 & 3.4   & 0.60  &  absorption\\
\coTwo\   &        &  0.15&0.11 & --80.9  && 0.7 & 0.5 & 1.7   & 0.40 &  absorption  \\
Continuum &  \phantom{+}87   &  0.16&0.09 & \phantom{+}58.1  && 0.8 & 0.4 &     & 0.06 &  \\
Continuum &  174   &  0.15&0.11 & --82.1  &  & 0.7 & 0.5 & & 0.10 &  \\
Continuum &  262   &  0.19&0.09 & --63.9  &  & 0.9 & 0.4 &  &0.11 &  \\
\hline
\end{tabular}
\tablefoot{ The maximum recoverable scales  of the \coOne, \coTwo\, and the \coThree\  observations are  4\farcs7 (22 kpc), 3\farcs8 (18 kpc) , and 2\farcs5 (12 kpc), respectively.}
\end{center}
\label{tab:obs}
\end{table*}

The properties of \target\ have been described in detail by \cite{Morganti21}. We summarise the main features below. 
\pks\ is a relatively young, powerful radio source ($\log P_{\rm 5\, GHz}/{\rm W\, Hz^{-1}} = 27.43$) that is hosted by an early-type galaxy (see \citealt{Ramos11} and references therein and Fig.\ \ref{fig:environment}), with a relatively high far-infrared luminosity for this type of galaxy \citep{Dicken09}. An accurate redshift of $z = 0.32188\pm0.00004$ was derived by \cite{Santoro20}\footnote{At this redshift, 1 arcsec corresponds to 4.716 kpc for the cosmology adopted in this paper, which assumes a flat Universe and the following parameters: $H_{\circ} = 70$ \kms\ Mpc$^{-1}$, $\Omega_\Lambda = 0.7$, and $\Omega_{\rm M} = 0.3$.} by fitting stellar absorption lines in a VLT/Xshooter spectrum, refining the estimate of  \cite{Holt08}.    The relatively young age of \pks\  was confirmed by the peaked radio spectrum shown in \cite{Tzioumis02} and \cite{Callingham17}. Interestingly, the latter found the spectrum to peak at a low frequency (i.e.\ 145~MHz) by fitting the radio spectral energy distribution. When we consider that the turnover frequency is predicted to shift to lower frequencies as the components of the source expand with time \citep{ODea98}, this indicates that the age of \pks\ is about  $ 10^5$ yr.

The radio continuum emission is distributed in two relatively symmetric lobes, as described by \cite{Tzioumis02}. The extent of the radio emission is about 3 kpc measured from peak to peak of the lobes,  and the full extent is $\sim$4.7 kpc. 
In addition to this, the high spatial resolution, combined with the high frequency of the ALMA 1.7~mm image of \cite{Morganti21}, enabled the detection of the core of \target, as shown in  Fig. \ref{fig:paper1}, reproduced from \cite{Morganti21}.
\target\ is also a strong optical AGN, with a bolometric luminosity in the range $L_{\rm bol}\sim 2.5$ -- $4 \times 10^{45}$ \ergs\  \citep{Santoro20} and an [OIII]$\lambda$5007 luminosity $L_{\rm [OIII]} = 1.5\times10^{42}$ erg s$^{-1}$ \citep{Dicken09}. This  qualifies it as a type 2 quasar according to the criteria of \citet{Zakamska06}. Strong and broad (${\rm FWHM}>1000$ \kms) optical emission lines were detected by \citet{Holt08}, \citet{Shih13} and \citet{Santoro20}. This large width indicates outflowing ionised gas 
with a mass outflow rate in the range $0.13<\dot{M}<1.47$ \msunyr. According to \citet{Santoro20}, the warm outflows have similar radial extents as the radio source, and they have a high density. The latter is explained  as the result of compression of the gas from the jet/ISM interaction. However, photoionisation (and not shocks) likely produces the emission lines (see \citealt{Santoro20} for details). 

Finally, three bright neighbouring galaxies that we identify in Fig. \ref{fig:environment} are confirmed to have similar redshifts as \target\ \citep{Tadhunter11}. The galaxy number counts  presented by \cite{Ramos13} also suggest a rich 
environment\footnote{The number count results for the environments of individual objects can be unreliable because they rely on an accurate subtraction of the
background galaxy counts, which can be challenging.}.  However, based on XMM observations, the total X-ray luminosity of \pks\ and its surroundings ($\log L_{\rm 2-10\,keV}/{\rm erg s^{-1}} = 43.27$; \citealt{Mingo14})  is lower than expected for a rich cluster of galaxies, and it is more consistent with that of a galaxy group \citep{Eckmiller11}.
The type of environment in which \target\ lives (whether cluster or group) is evidently still unclear, and this has motivated our new \chandra\ X-ray observations.    These observations show that PKS 0023--26 is not part of a galaxy cluster or rich group, and it is at most a member of a poor, low-temperature ($kT \lesssim$ 0.5 keV) galaxy group. A detailed analysis is  presented in the companion  paper.

In this paper, we aim to further investigate the proposed scenario of the interplay between the AGN in \target\ and the rich medium in which it is embedded based on new ALMA observations.  
The paper is organised in the following way. The new ALMA \coOne\ and \coThree\ observations are presented in Sect. \ref{sec:observations}.
The results for the molecular gas, obtained by combining \coOne\ and \coThree\ with the available \coTwo\ data, are presented in Sect. \ref{sec:molecular}. 
In Sect.\ \ref{sec:discussion} we combine our findings and describe the energetics and the proposed scenario, also by comparing \target\ with BCG and other AGN.

\section{ALMA observations of \coOne\ and \coThree}
\label{sec:observations}

The parameters of the ALMA {   CO(1-0) and CO(3-2)} observations that were performed during Cycle 9 are presented in Table \ref{tab:obs}    and were chosen to match those of the CO(2-1) observations of \cite{Morganti21}, the details of which are given in this table as well.
This allowed us to study the distribution of the molecular gas on similar spatial scales as the extended radio emission (about 1 arcsec {   or $\sim$ 5 kpc}).

   \begin{figure*}[h]
   \centering
      \includegraphics[angle=0,width=\hsize]{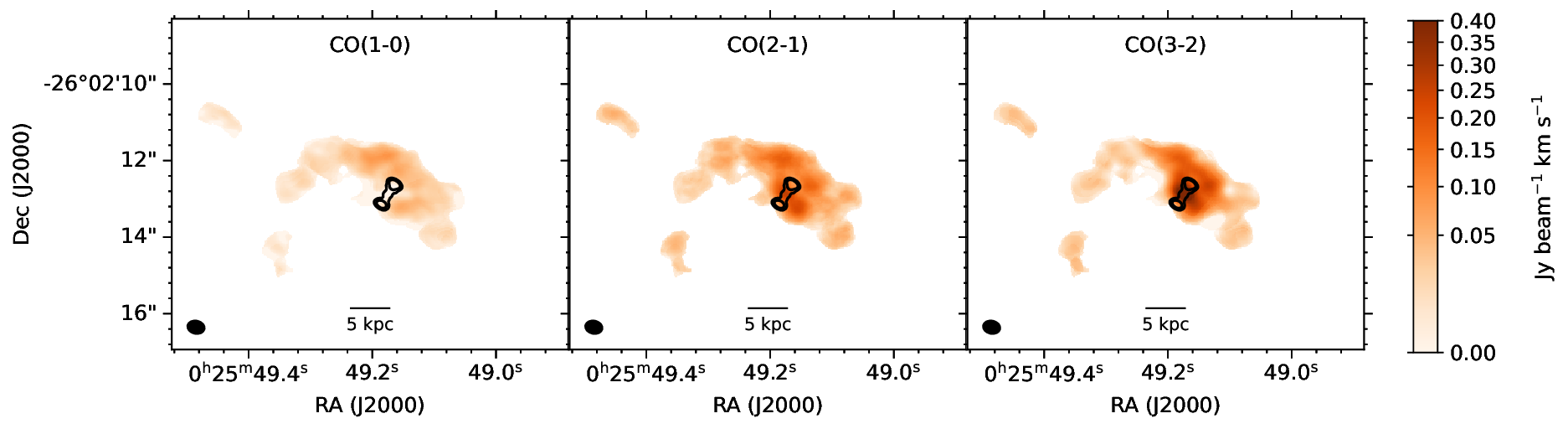}
   \caption{ Total intensity images of the three transitions on the same flux scale (Jy beam$^{-1}$ \kms).    These images were made from data cubes convolved to the same resolution of 0\farcs45 $\times$ 0\farcs34  (2.1 $\times$ 1.6 kpc; see Sect.\ 2). In all three panels, the 87-GHz continuum is overplotted with contour levels at 1 and 9 \mJybeam. 
 }
              \label{fig:molecularGas}
    \end{figure*}

The ALMA CO(1-0) observations  were obtained in configuration C43-7 in four observing sessions, two on June 11 and two on June 12, 2023, resulting in a total on-source observing time of 46.32 min.  
The observations were made in Band 3 and used the correlator in frequency-division mode. One spectral window was centred on the redshifted \coOne\ emission using a central sky frequency of 87.195 GHz and a total bandwidth of 1.875 GHz (corresponding to 6450 \kms). With the 1920 channels we used, the native velocity resolution is 3.4 \kms\, but the data were binned to a   lower spectral  resolution  when we made the final data cube (see below). The maximum recoverable scale of these observation is 4\farcs7    (22.2 kpc).

The \coThree\ observations were obtained using ALMA in configuration C43-5 in  one observing session on April 22, 2023, with a total on-source time of 83.18 min.   
The observations were made in Band 6 with the correlator in frequency-division mode. One spectral window was centred on the  \coThree\ emission using a central sky frequency 262.273 GHz and a total bandwidth of 1.875 GHz (corresponding to 2150.0 \kms). With the 1920 channels, the native velocity resolution is 1.1 \kms, but in the subsequent data reduction, the velocity channels were  binned to match the spectral resolution of the CO(1-0) and CO(2-1) cubes   (see below). The maximum recoverable scale of these observation is 2\farcs5  (11.8 kpc).
The source J2253+1608 was used as bandpass, flux, and atmospheric calibrator for the observations of both transitions. J0011--2612 was used as phase calibrator for the \coOne\ observations and  J0038--2459 for the \coThree\ observations.    In general, the main uncertainty in the derived ALMA fluxes comes from the limited accuracy of the flux calibration, which
is estimated to be about 5\% \citep{vanKempen14}. The data in the ALMA calibrator database show, however, that the flux of J2253+1608 varied rapidly during the period of our observations. To be on the safe side,  we therefore assumed that the error on the derived fluxes is 10\%.

The  calibration was made in CASA (v6.4.1.12; \citealt{McMullin07}) using the  reduction scripts provided by the ALMA observatory. 
The visibility products  resulting from the ALMA pipeline were found to be of sufficient quality, and no further calibration was made. The calibrated $uv$ data produced by the pipeline were exported to MIRIAD \citep{Sault95} for the further analysis.

For both the \coOne\ and \coThree\ line observations, data cubes were made with the same  velocity resolution as the existing \coTwo\ data cube (33.6\kms)    by binning the channels to 16.8 \kms\ and applying Hanning smoothing. We made them using natural weighting.  The binning improves the S/N of the emission in the channel maps while still  resolving the individual spectra, the narrowest of which have a velocity dispersion of about 25 \kms. Before making these cubes, we removed the absorption seen in the \coOne\ data against the northern lobe (see below). The continuum was subtracted in the image domain by fitting for every position a straight line through the line-free channels and subtracting this fit from the observed spectrum. The channel images were cleaned down to half the noise level  using  masks that were made by smoothing the cube to twice the spatial resolution and using the 2$\sigma$ level of this smoothed cube to create the mask for the original cube. After cleaning, the \coOne\ and \coThree\ data cubes were de-redshifted using $z = 0.32188$. 

As mentioned above, strong absorption was  detected against the N lobe in the \coOne\ data cube, but not in the CO(3-2) cube. This complements the absorption that was earlier also detected in \coTwo\ \citep{Morganti21}. To study this absorption, we also made \coOne\ and \coTwo\  data cubes  at the original velocity resolution of the observations (3.4 \kms\ and 1.7 \kms\, respectively{; see Table \ref{tab:obs}}).  For these cubes, we  used uniform weighting, and thus, a higher spatial resolution, in order to minimise the effect of confusion of the absorption spectra with emission. The properties of this absorption  are discussed in Appendix \ref{sec:Abs-Appendix}. We also used these cubes with the high velocity and spatial resolution to remove the absorption from the cubes with a lower velocity resolution, however. Because the surface brightness of the CO emission is relatively low,  the high spatial resolution  and the higher noise level of these cubes means that most of the CO emission is   below the noise level, while the absorption is detected at high significance (see Fig.\ \ref{fig:absorption}). In order to obtain a model of the absorption, we therefore only used  the deconvolution task clean  on these high-resolution cubes at the location where absorption was seen and subtracted these   negative clean components from the original $uv$ data sets before we made the low-velocity resolution cubes. 

The total intensity (i.e.\ moment-0) images of \coOne\ and \coThree\ are shown in Fig.\ \ref{fig:molecularGas} (left and right image, respectively), together with the image of the  \coTwo\ emission (centre) from \cite{Morganti21} for comparison.  Because we wished to use these images to make images of the line ratios, we convolved the CO(1-0) and CO(3-2) cubes to the same spatial resolution as the \coTwo\ cube  (0\farcs45 $\times$ 0\farcs34 $\equiv 2.1 \times 1.6$ kpc) and regridded them to the same spatial and velocity grids as the \coTwo\ observations.  In these cubes, the peak S/N of the emission is 12 $\sigma$ in the CO(1-0) cube and 18 $\sigma$ in both the CO(2-1) and CO(3-2) cubes.

The moment-0 images were obtained by first running the SoFia-2 source-finding software \citep{Westmeier21} on the \coTwo\ data cube to construct the moment images for this line transition, and we subsequently applied the same mask generated by Sofia-2 on the cubes of the two other transitions to make the moment images for these transitions as well because the typical S/N level of the CO emission is quite different for the three transitions.  Therefore, using masks that were derived separately for the three transitions would mean that the spectra of the different transitions are integrated over different velocity  ranges, which would lead to S/N-dependent biases in the line ratio. We chose the \coTwo\ cube as the benchmark because the typical S/N of the emission is  highest in this data cube.

Continuum images  were made from the line-free channels, and the parameters of these images are summarised in Table \ref{tab:obs}. More details on the continuum are given in Appendix \ref{sec:AppendixCont}.  

   \begin{figure}
   \centering
   \includegraphics[angle=0,width=8.5cm]{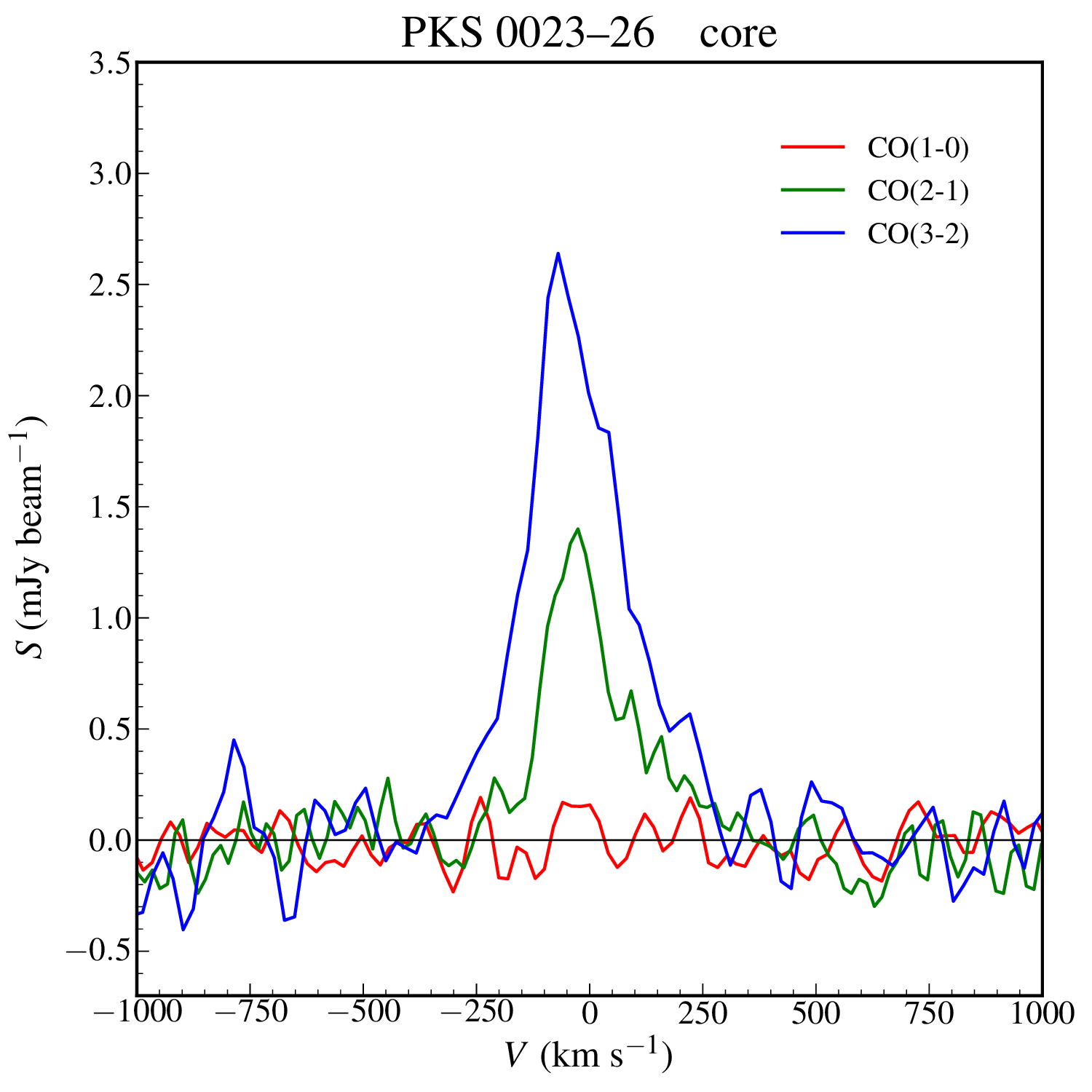}
   \caption{Spectra of the three transitions at the location of the radio core.  
 }
              \label{fig:ProfilesCore}
    \end{figure}

   \begin{figure*}
   \centering
      \includegraphics[angle=0,width=17cm]{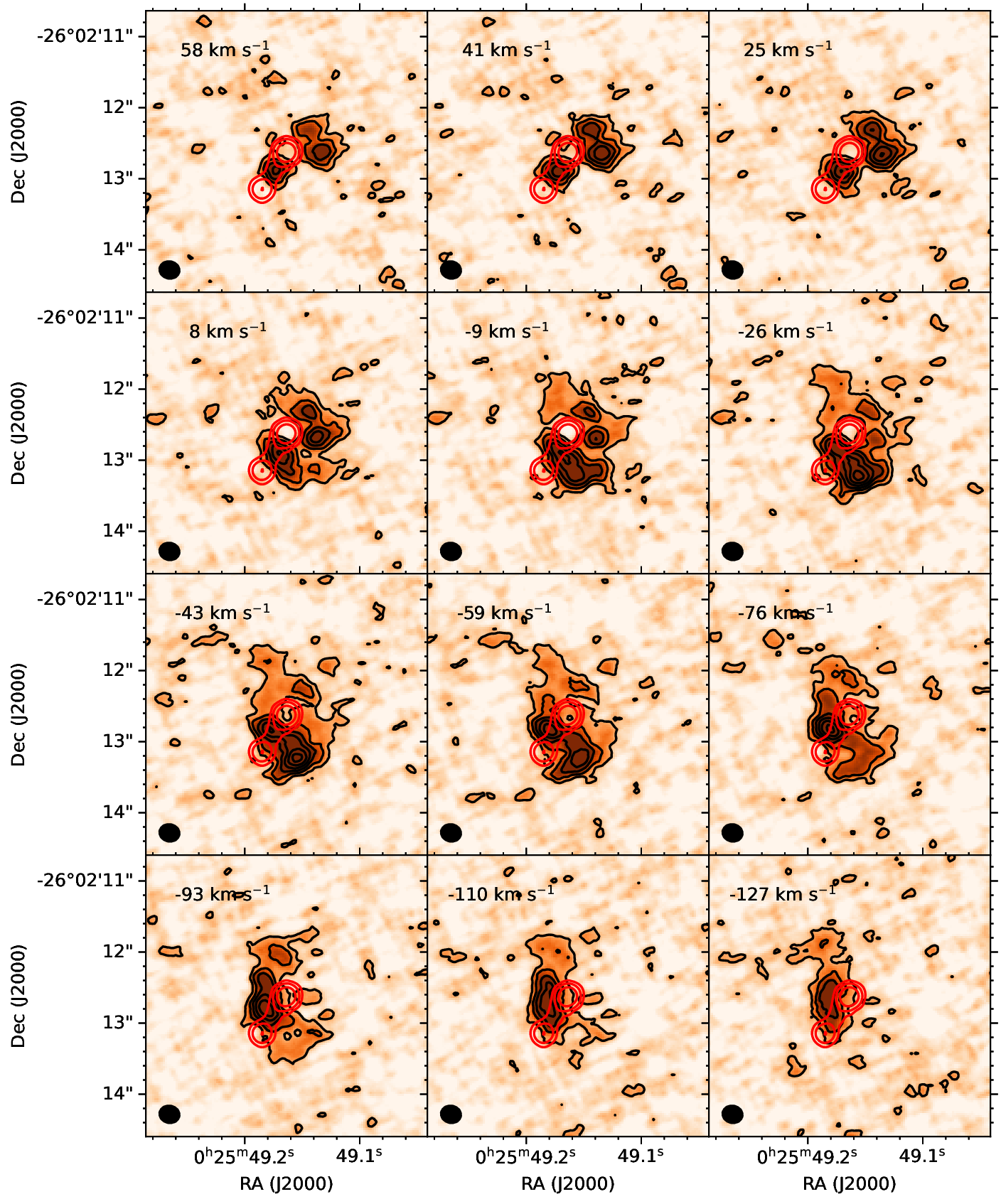}
   \caption{Subset of channel maps to illustrate the {   broad velocities in the centre and  the \coThree\ emission  wrapping around the northern lobe (contours in red at levels 1, 3, and 9 \mJybeam). }
   }
              \label{fig:channels}
    \end{figure*}

\section{Results: The molecular gas}
\label{sec:molecular}

The moment-0 images presented in Fig.\ \ref{fig:molecularGas} show that the addition  of  observations of two more CO transitions provides very useful information on what occurs in \pks. 
We first discuss the results for the distribution  of the gas that is observed in emission (Sect. \ref{sec:COdistribution}). 
We then discuss the kinematics of the gas in Sect. \ref{sec:COkinematics}, and we study the line ratios in Sect.\ \ref{sec:lineRatios}.

\begin{figure*}
   \centering
      \includegraphics[angle=0,height=7cm]{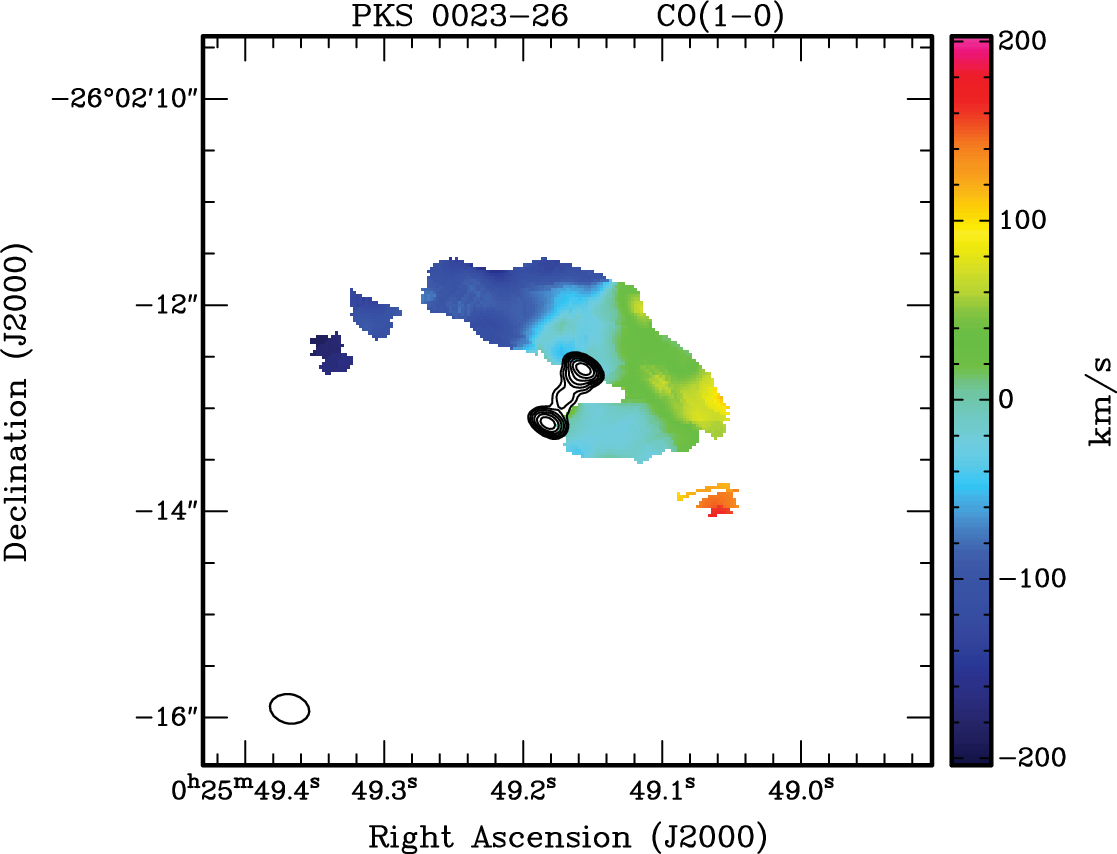}
      \includegraphics[angle=0,height=7cm]{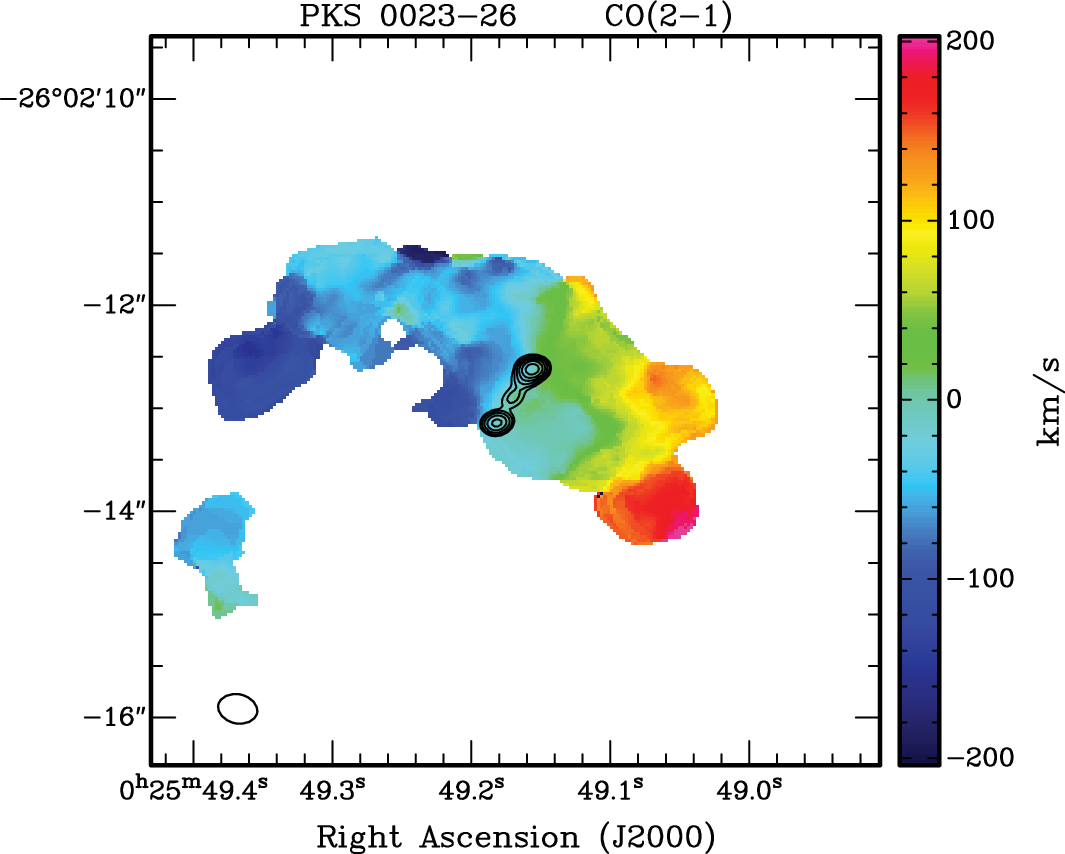}
      \vskip5mm
      \includegraphics[angle=0,height=7cm]{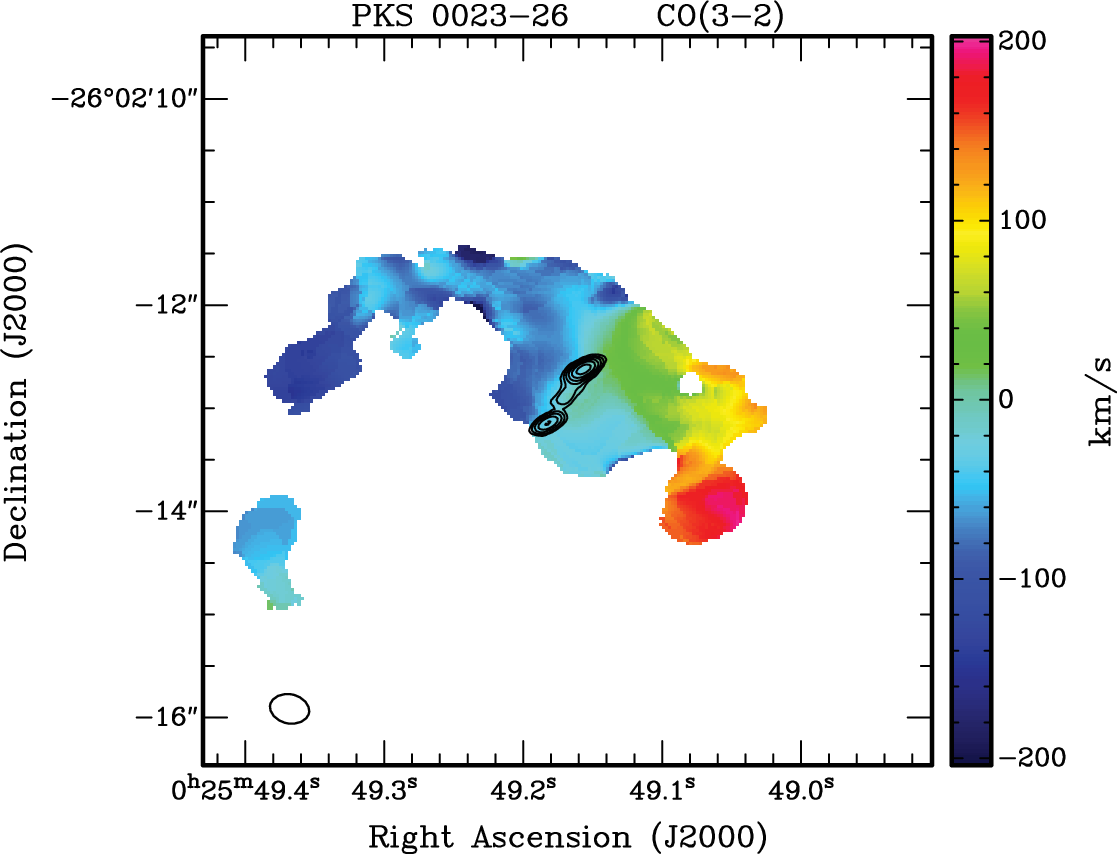}
      \includegraphics[angle=0,height=7cm]{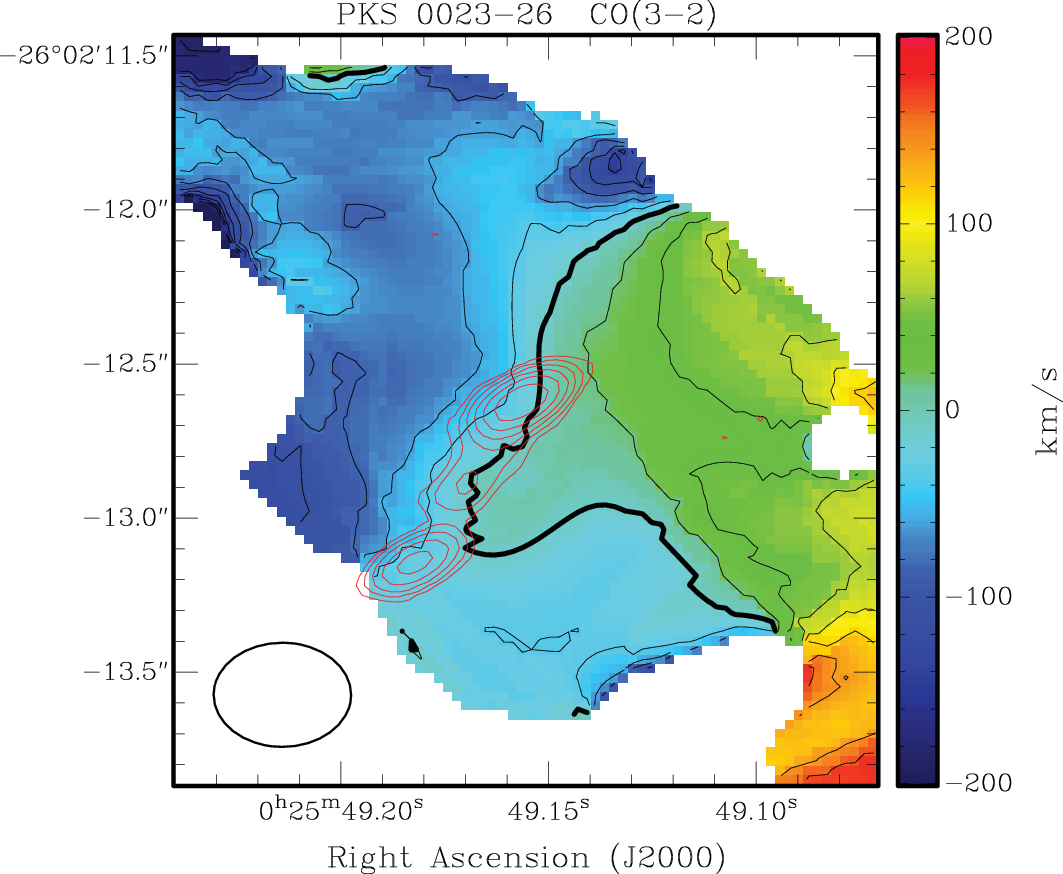}
   \caption{ {\sl Top row and bottom row left}: Velocity fields of  the three CO transitions . The velocities are only shown for locations in which the S/N of the integrated emission is higher than 5. The contours show the continuum emission at the same frequency as the line emission of that particular panel (contour levels 1, 2, 4, ... \mJybeam).    {\sl Bottom row, right:} Zoom-in of the CO(3-2) velocity field with iso-velocity contours (--200, --175, ..., +200 km/s). The systemic velocity is indicated by the thick contour. The 262-GHz continuum is indicated by the red contour (levels 0.4, 0.8, 1.6, ..., 12.8 \mJybeam). The beam size is indicated in the bottom left corner of each panel. 
   }
   \label{fig:velfies} 
\end{figure*}

\subsection{Distribution of molecular gas in the three CO transitions}
\label{sec:COdistribution}

The extent and morphology of the distribution of the  \coThree\  is similar to that of the \coTwo\  \citep{Morganti21}, although clear differences can  be noted in the relative intensity of these two transitions at different locations. This suggests different conditions of the gas (Fig.\ \ref{fig:molecularGas}). This is even more extreme in the case of \coOne. Overall, the  molecular gas is distributed over a region with a diameter of about 3--4 arcsec (14--19 kpc) that surrounds the radio source,  but is not centred on it,  and is more extended  to the W and N. In particular, while the northern lobe appears to be embedded in the molecular gas, the southern lobe is not.

From the main body around the radio source, a tail or tidal feature extends toward a nearby galaxy $\sim$3 arcsec ($\sim$14 kpc) east of \target\ (A in Fig.\ \ref{fig:environment}). A tail in the  SW direction is also seen, although less clearly, and possibly pointing in the direction of another companion galaxy (B or C2). A small cloud is seen close to another companion (C1), but it is offset from its centre and has an  elongated morphology roughly in the direction of \target. A more detailed discussion of the environment and origin of the gas will be given in the companion paper \citep{Siemiginowska2025}.

The brightest region of \coThree\ emission is (similar to what is seen in \coTwo\, but more clearly)  seen at the location of the radio core and extends in a direction roughly perpendicular to the jet (Fig.\ \ref{fig:molecularGas}). The diameter of this structure is $\sim$0.8 arcsec ($\sim$3.7 kpc) from one side of the nucleus to the other. 
Interestingly, and in contrast to what we observed in \coTwo\ and \coThree,  this region appears to be depleted of \coOne\ gas. This is illustrated in  Fig.\ \ref{fig:ProfilesCore}, where the profiles of the three transitions taken at the position of the core of \pks\ are shown. At this location, no \coOne\ emission is found, but the other  transitions are clearly detected. This is clearly different in the outer regions, where the \coTwo\ and \coThree\ are fainter than near the core, but where the \coOne\ is detected instead. This appears to be similar to the situation in  NGC~3100, where a central lack of CO(1-0) is observed \citep{Ruffa22}, or in ESO 428--G014, where a lack of CO(2-1) is seen in the central regions while H$_2$ gas is clearly detected \citep{Feruglio20}. Feruglio et al. (2020) suggested that X-ray photons in these regions might excite CO to higher excitation levels, which would explain the low levels of the low-$J$ CO emission. Something similar may also occur in NGC 2110 \citep{Fabbiano19}. We discuss this further in Sect.\ \ref{sec:lineRatios}.

In the region of the radio lobes, the CO appears to be faint  in \coTwo\ and \coThree. This is clearly visible in the channel maps in Fig.\ \ref{fig:channels}, where the distribution of the \coThree\ gas (grey scale, black contours) for a range of velocities (from $+58$ to $-128$ \kms)  is shown. The continuum emission (at  262 GHz) of the radio galaxy is superposed  in red contours. The CO emission clearly avoids the locations with the strongest continuum emission. This was already noted by \cite{Morganti21} for the \coTwo\ emission. For this transition, we were unable to completely exclude some effect of the CO absorption seen against the northern lobe (even though \cite{Morganti21}  removed this absorption from the CO(2-1) data cube). In contrast to \coOne\ and \coTwo, no \coThree\ absorption is detected against the northern lobe, and this is therefore no problem here.

   \begin{figure}
   \centering
   \includegraphics[angle=0,width=8.5cm]{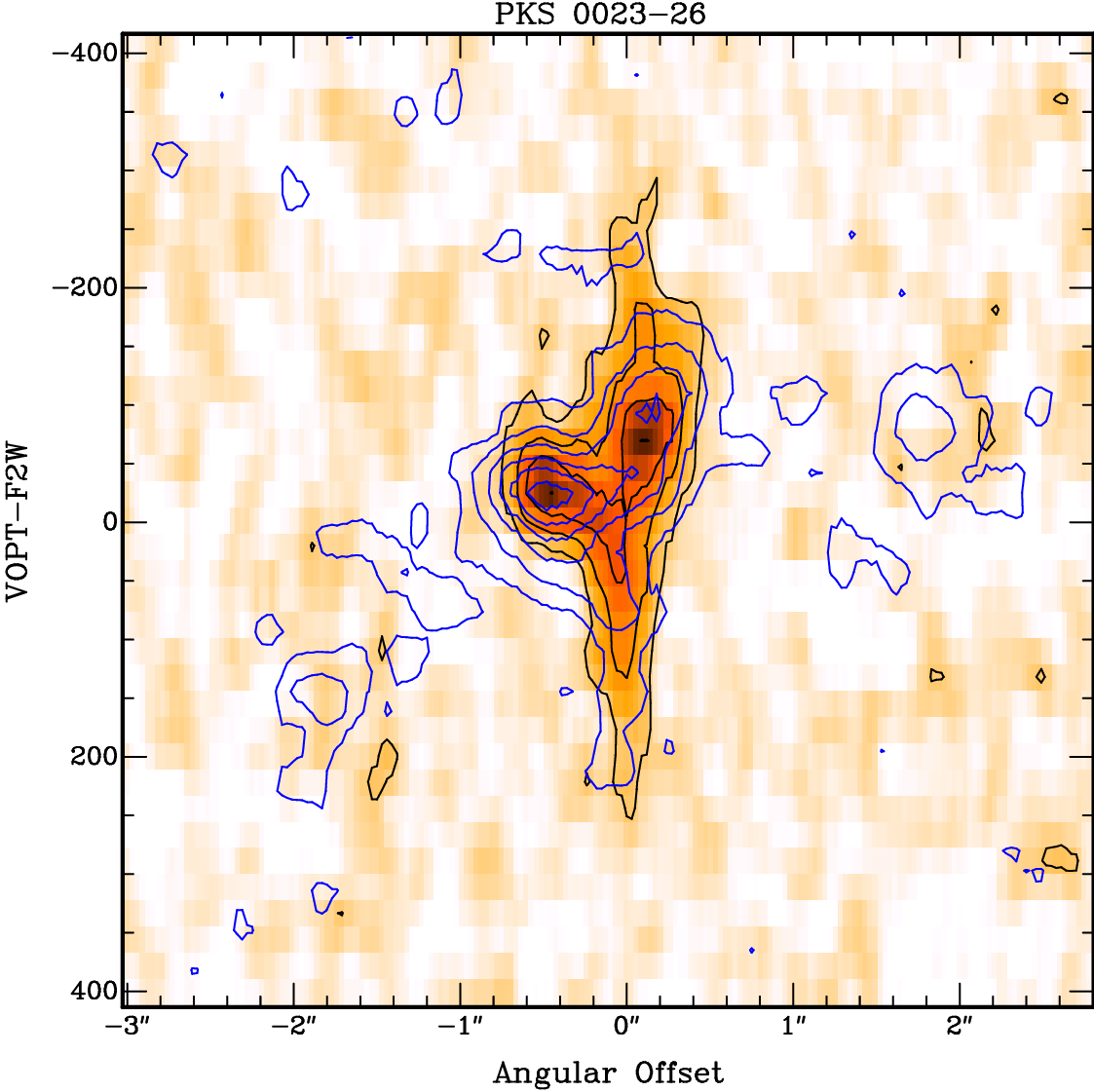}
   \caption{Position-velocity plot of the  \coThree\ emission (grey scale) obtained from a cut  centred at the core and perpendicular to the radio axis (PA $= 51^{\circ}$). The plot shows the  large velocity width (FWZI $\sim$ 600 \kms) of the gas in the central regions. The contour levels are at 0.36, 0.72, and 1.08 \mJybeam. The CO(2-1) emission along the same position angle (blue contours, levels at 0.24 and 0.48 \mJybeam) shows the same broad emission in the centre, but also the overall velocity gradient in this direction.
  }
   \label{fig:PositionVelocity}
 \end{figure}

\subsection{The complex kinematics from small to large scales}
\label{sec:COkinematics}

The overall velocity field observed  for the three transitions is shown in Fig.\ \ref{fig:velfies} , where  a zoom-in of the velocity field of the CO(3-2) is also shown. The \coOne\ and \coThree\ show the same smooth velocity gradient that was already observed for the \coTwo. The kinematics of the gas at the location and around the radio source is particularly interesting, however.

Since no \coOne\ is detected in the central regions, we used the \coThree\  to investigate the kinematics of the gas.    The zoom-in velocity field of the CO(3-2) clearly shows that while  there is an overall SW-NE gradient, the   large-scale kinematics does not correspond to that of a regularly rotating disk.
As mentioned above, the most extreme kinematics of the gas is  observed in the central region. Figure \ref{fig:PositionVelocity} shows the position-velocity plot   of the CO(3-2) emission  at a  position angle $51^{\circ}$ (i.e.\ perpendicular to the radio axis), which illustrates the large   profile width in the central regions, with a full width zero-intensity of FWZI $\sim$600 \kms\ (FWHM $\sim$300 \kms),  
consistent with the profiles shown in Fig\ \ref{fig:ProfilesCore}. Figure \ref{fig:PositionVelocity} also shows that the  kinematics of the gas near the nucleus does not seem to be consistent with a regularly rotating disk,  and its kinematics clearly differs from the overall large-scale gradient in velocity. This suggests that the broad profiles are driven by the AGN.
Although a broad, central  profile was already seen in  the \coTwo\ data, the new \coThree\ observations show a more symmetric profile with a broader blueshifted component than is observed in \coTwo\ \citep{Morganti21}. The reason is probably that the \coThree\ is brighter than the \coTwo\ in this region and has a higher S/N (see Fig.\ \ref{fig:ProfilesCore} and Sect.\ \ref{sec:lineRatios}). 
In this direction perpendicular to the radio jet, the broad component appears to be spatially limited to one beam, and it therefore is on a scale of an  FWHM $\sim$1 kpc. 

\begin{figure*}
  \centering
    \includegraphics[angle=0,width=9cm]{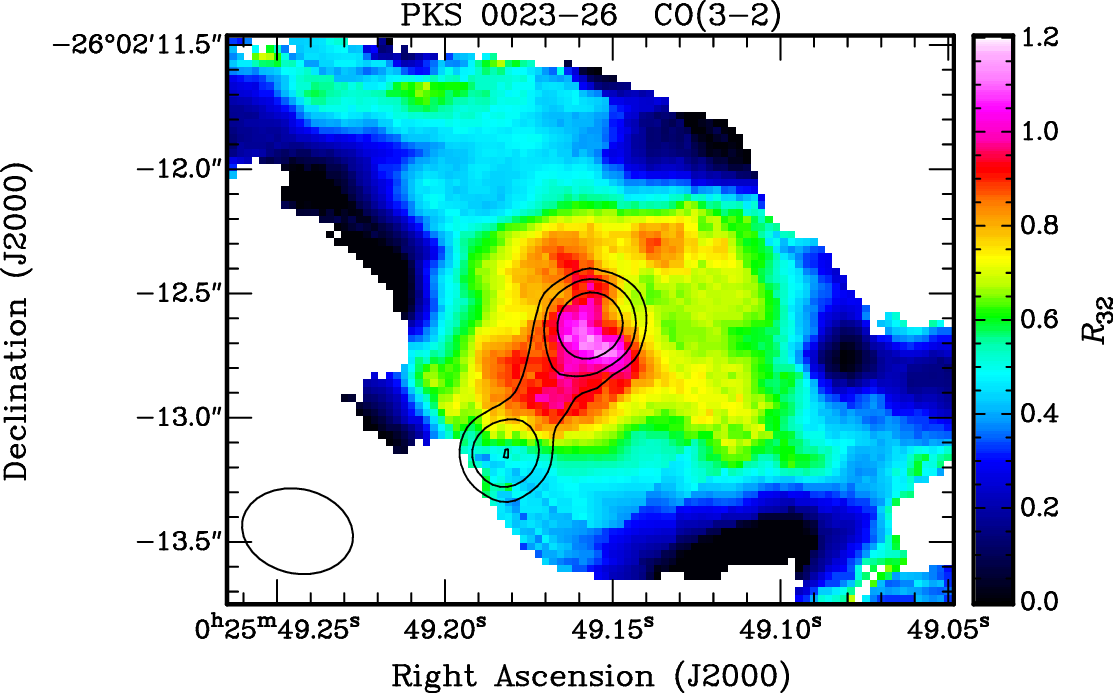}
    \includegraphics[angle=0,width=9cm]{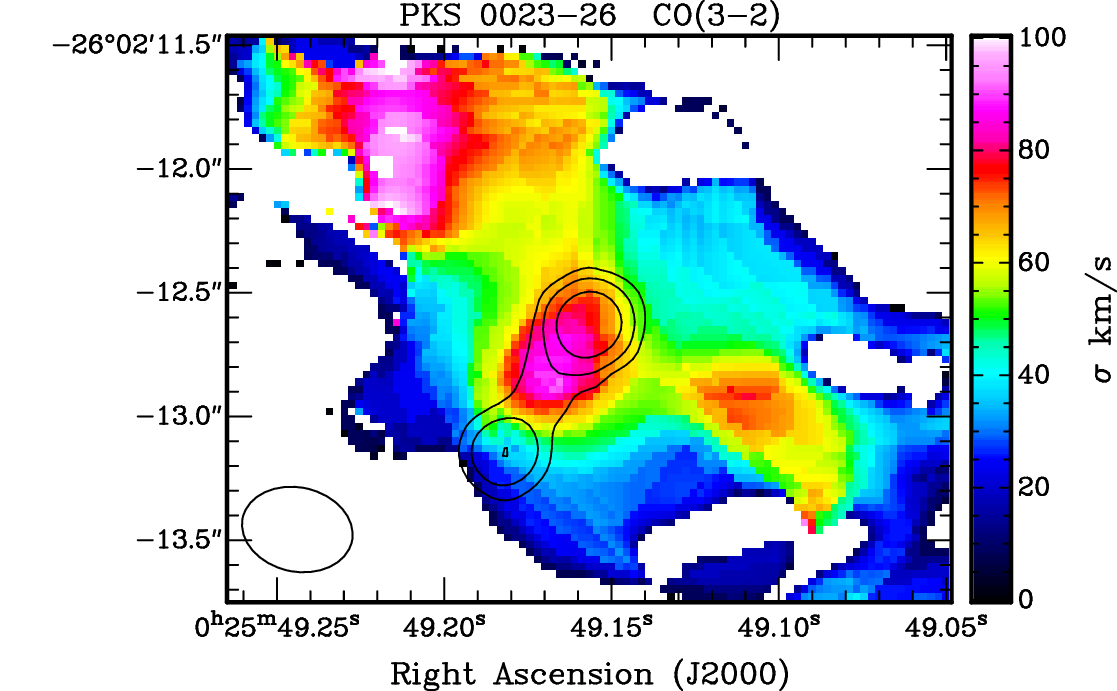}

\caption{   
{\sl Left:} Velocity dispersion of the \coThree\ emission. The 87 GHz radio continuum contours are superposed.
{\sl Right:} Brightness temperature ratio $R_{\rm 32}\equiv I_{\rm 3-2}/I_{\rm 2-1}$ . The contours of the radio continuum are superposed. The contour values for the radio continuum are 1, 2, 4, 8 ... \mJybeam\ in both panels. The beam size in indicated in the bottom left corner of each panel.}
\label{fig:LineRatio}
\end{figure*}

As mentioned above, the CO emission appears to avoid the northern lobe. The emission at this location is not only very faint (see Fig.\ \ref{fig:channels}), but also  relatively broad in velocity. This can be seen in  the velocity dispersion ($\sigma$) image presented in Fig.\ \ref{fig:LineRatio}, which shows that 
in the region between  the core to the northern lobe, $\sigma$ reaches almost 100 \kms,  which is well above the typical values for   undisturbed  molecular gas in normal galaxies ($\sigma \sim 10 - 20$ \kms; \citealt{Ruffa24}).  Just beyond the radio source, the velocity dispersion is still about 50 \kms. 

As mentioned above, on the largest scales, the \coOne\ and \coThree\  show the same smooth velocity gradient as was observed for the \coTwo\ (Fig.\ \ref{fig:velfies}). In the north-eastern part of this extended gas distribution, a region with a high velocity dispersion is found. As remarked in \cite{Morganti21}, this is likely to be due to the superposition of two gas structures, as is evident from double-peaked profiles at these locations. About 1 arcsecond south-west of the radio source, another region has large velocity dispersions. This is likely due to beam smearing, given the large, local velocity gradients at this location.

\noindent

\subsection{Line ratios}
\label{sec:lineRatios}

The total intensity images in Fig.\ \ref{fig:molecularGas} clearly show that  the relative intensity of the three  different transitions strongly changes with position. This is particularly extreme in the central beam, where the \coOne\ is undetected, while the other two transitions are relatively bright. In the outer regions, the  contrast is much lower. To investigate this in more detail, we constructed line-ratio images of the observed CO transitions.

The plot in Fig.\ \ref{fig:LineRatio} shows the distribution of the brightness temperature ratio $R_{\rm 32}\equiv I_{\rm 3-2}/I_{\rm 2-1}$, and the 87-GHz radio continuum (in grey contours) is superposed for reference. This image clearly shows that \rdt\ peaks in the region between the core and the northern lobe. The values are in the range of 1.0 to 1.2. No high values like this are seen at the location of the southern lobe, where we also observed a lack of gas (see above). 

Despite the depth of the observations, the region of the radio source of \target and slightly beyond is depleted of \coOne. We estimated an upper limit to the CO(1-0) flux by assuming the same line width as for CO(3-2), and this gives a 3$\sigma$ upper limit to the \coOne\ flux integral of the central region of 39 mJy \kms.
This means  a lower limit of $R_{21} \gtrsim 1.9$, which is quite extreme. At the locations where  $R_{32} < 0.6$, CO(1-0) is detected, and values more typical for a normal ISM are seen ($0.4 < R_{21} < 0.6$).

For comparison, the molecular gas in star-forming galaxies shows typical values of \rto\ around 0.6, with a standard deviation of about 0.2 \citep{Brok21,Yajima21,Leroy22}, and  \rdt\ ranges between 0.2 and 0.6 (16\% -- 84\% range; \citealt{Leroy22}). For early-type galaxies, similar values for \rto\ are found from spatially resolved observations (e.g.\ \citealt{Topal16,Young22} and Davis priv. comm.). Values like this have also been found in high-$z$ galaxies (e.g.\ \citealt{Kaur25}). 
On the other hand, a broad distribution of $R_{\rm 32}$ and $R_{\rm 21}$ values is seen in ULIRG/LIRG objects, including ratios higher than one (e.g.\ \citealt{Montoya23}). However, these studies reported integrated values and did not resolve the distribution of the gas, as we did for \target. Thus, it is difficult to identify whether the high ratio is connected to particular regions in ULIRG systems, specifically, to regions that might be affected by the AGN (when present). For example, in the case of the LIRG NGC~6240, the region of higher line ratios ($R_{\rm 21} > 1$ ) is associated with outflowing gas, and in general, with higher velocity dispersion components. This region appears to be located between the two AGN \citep{Cicone18}. For another ULIRG, however, IRAS 10190+1322, \cite{Gracia07} found a brightness temperature ratio $R_{\rm 21}$ between 0.6 and 0.78, which is at the upper end of what is found in normal spiral galaxies.

Figure \ref{fig:LineRatio} shows that the region with elevated line ratios extends well outside the radio source.  A region extending up to $\sim$5 kpc from the radio source shows ratios higher than 0.6. This might indicate that a large region might be directly or indirectly affected by the AGN, and in particular, the radio plasma, in regions that lie well outside the radio source.  In these regions, however, we cannot exclude the possibility that other mechanisms that are connected to the on-going galaxy interaction also affect the gas conditions (see Sect. \ref{sec:interaction}).

Although there is no exact one-to-one correlation,  the distribution of  $R_{\rm 32}$ is similar to the structure of the velocity dispersion shown in Fig.\ \ref{fig:LineRatio}. In both cases, the regions with the strongest enhancements appear to be aligned with
the radio axis on the NW side of the nucleus. In addition to the central region with high velocity dispersions and high values for \rdt, both extend to the west, where high values are  found.
Structures with high line ratios and/or a high velocity dispersion that extends perpendicular to the radio axis are seen in a growing number of objects (e.g.\ \citealt{Audibert23,Murthy25}), but they are smaller in scale than in \target.

Finally, the results for the excitation of the gas allowed us to re-estimate the total mass of the molecular gas by narrowing down the parameters used by \cite{Morganti21}.  The combination of conversion factors that seems to be most appropriate considering the observed line ratio gives an H$_2$ mass of  $M_{\rm mol} = 6.2 \times 10^{9}$ \msun\ for a conversion factor $\alpha_{\rm CO}$  typical of ULIRGs, $\alpha_{\rm CO} = 0.89  \ M_\odot/({\rm K\ km\ s^{-1}\ pc^2}) $ and $R_{21}=1$. The latter represents a compromise between the value measured at large radii and the value in the central part. When we instead assumed that all the gas had a typical Galactic  $\alpha_{\rm CO} = 4.3  \ M_\odot/({\rm K\ km\ s^{-1}\ pc^2}) $, this gave an upper limit of the molecular mass of $M_{\rm mol} = 3.1 \times 10^{10}$ \msun\ (see also \citealt{Morganti21}).

\section{Discussion: AGN feedback and evolution of the radio jets}
\label{sec:discussion}

We have traced the distribution and kinematics of the cold molecular gas, and we mapped the diversity of the physical conditions of this gas in \pks\ using different transitions. The observations confirmed that the AGN has a significant impact on the ISM and added evidence of how this impact appears to evolve as the radio jet expands. 

\subsection{Jet-ISM interactions}
\label{sec:interaction}

The new ALMA observations show several signatures of not only the impact of the AGN, but more specifically, of the on-going interaction between the jet and the ISM. We can even trace and quantify the impact. In the central kpc, we found broad line profiles and disturbed kinematics for the molecular gas. This was already seen in \coTwo\, but appears even more prominently in the new \coThree\ observations.
The broad range in velocities that is observed (FWZI$ \sim$ 600 \kms) suggests that the gas is likely associated with an outflow driven by the AGN. This was also observed for the warm ionised gas by \cite{Santoro20}. The extreme conditions of the gas in the central regions are further illustrated by the values of the line ratios of the various CO transitions   and by the non-detection of CO(1-0) in the central regions. The size of the region in which the outflow of cold molecular gas is observed ($\sim$1 kpc) is similar to that seen in other objects (as summarised in Sect.\ \ref{sec:introduction}). In some cases that were presented in the literature (e.g.\ \citealt{Dasyra16,Oosterloo17,Ruffa22,Murthy22,Murthy25}),  the morphological association between the extended radio continuum and the region in which the outflow occurs, or in some cases, the absence of a powerful optical AGN, suggests that the radio jet probably produces the outflows.  In \pks, the near-nuclear outflow is spatially unresolved in our observations, but the galaxy also hosts an AGN with a high bolometric luminosity. Thus, it is not possible to determine whether the radio jet or a radiation-pressure-driven wind accelerates  this compact outflow.

The distribution, kinematics, and line ratios  of the molecular gas on larger scales, however, suggest that the jet interacts with the ISM  on larger scales, especially at the location of the N lobe and in the regions between this lobe and the nucleus. 
At the position of the N lobe, there is a lack of molecular gas, while molecular gas appears to be wrapped around the lobe at larger distances (Fig.\ \ref{fig:channels}). 
The former is consistent with a strong interaction between the expanding radio jet/lobe and the molecular ISM, which probably causes the gas clouds to be shredded and heated by hydrodynamical instabilities as they interact with the  hot wind behind the jet-induced shock \citep[e.g.][]{Klein94,Mellema02}.  If the pre-existing warm/cold gas had fairly low  densities (perhaps not unlikely to be the case at the radius of the N lobe), it may not have survived the passage of the jet and cooled back
to a molecular phase. This would  explain the lack of molecular gas at the sites of the lobes. There is evidence in some local radio galaxies that strong 
jet-cloud interactions can indeed be so destructive that the warm/cold precursor gas does not cool behind the jet-induced shock. PKS~2152--69 and 3C~277.3 (Coma~A) both show strong jet-cloud interactions in which the jets are deflected in the interactions, but the warm gas kinematics of the interacting gas clouds show only mild (PKS~2152--69; \citealt{Tadhunter88}) or no (3C~277.3; \citealt{Inarrea03}) signs of kinematic disturbance at radio knots that mark the locations in which the jets are deflected.  In the case of \target, direct evidence for the destruction of gas clouds at the site of the N lobe may be provided by the detection of enhanced soft X-ray emission at this location in our deep Chandra observations, as reported in the companion paper. These observations reveal  hot gas that is aligned with  the radio source. The X-rays are enhanced at the N radio lobe and at the location of the peak of $R_{32}$, suggesting that the X-ray emission could play a role in the excitation of CO. 
Evidence that the radio jet can drive shocks in the ISM and produce the X-ray emission was seen in other cases, for example,   NGC~4151 \citep{Wang11}, Mkn~573 \citep{Paggi12}, and 3C~305 \citep{Hardcastle12} (see also \citealt{Fabbiano22} for an overview). In \pks\ we detect some molecular gas even in the extended region between the core and the N lobe, however, which is characterised by an enhanced velocity dispersion ($\sigma\sim$ 80 -- 100 km s$^{-1}$) and very high CO excitation ($R_{32} > 0.9$) compared with its immediate surroundings (see Fig.\ \ref{fig:LineRatio}). 
Thus, even if not all the molecular clouds are destroyed by the interaction, the physical properties of the gas are profoundly affected by it. 

At larger distances from the core, the molecular gas tends to wrap itself around the N radio lobe (Fig.\ \ref{fig:channels}).  These regions are still characterised by values of $R_{32}$ that are higher than the values expected for an ISM like this (i.e.\ $R_{32} > 0.6$).
Moreover, the region with these high line ratios extends a few kpc beyond the N radio lobe to the north-west and to the west of the nucleus, in a direction that is more or less perpendicular to the radio axis. This might suggest that a milder interaction occurs there that might be created by the more slowly expanding cocoon around the radio source. This is also suggested by the lack of strong kinematical disturbances at these locations, except for the velocity dispersion (observed to be in the range 40-50\kms), which is higher than what is typically found in an  undisturbed ISM.

Overall, the values of $R_{\rm 21}$ and $R_{\rm 32}$ measured in and around \pks\ clearly deviate from the ratios found in the    undisturbed ISM in normal star-forming galaxies. We observe $R_{21} > 1.9$ and $R_{\rm 32}> 0.9$ in the region of the radio source, while over a more extended area, \rdt\ is still well above 0.6 (Fig.\ \ref{fig:LineRatio}) and is thus very different from what is expected from an   undisturbed ISM in normal galaxies. 
The coincidence of the most extreme line ratios with the region of the radio emission strongly suggests the impact of the radio jet. 

As mentioned above, a high excitation of the molecular gas is a common feature observed in radio AGN when multiple CO transitions are available.  The best-studied case is IC~5063 \citep{Dasyra16,Oosterloo17}, which is one of the best examples of a jet that affects the surrounding medium, as confirmed by the kinematics of various phases of the gas (atomic, molecular, and ionised, and using the same CO transitions as for \target), as well as by numerical simulations \citep{Mukherjee18b}. 
In IC~5063, the cold molecular gas associated with the outflow and the jet-ISM interaction is much brighter in \coTwo\ and \coThree\ than in \coOne\ and  has values between 0.97 and 1.25 for $R_{\rm 32}$ and between 1.06 and 1.79  for $R_{\rm 21}$. These values are similar to what is seen in \target.  For the fastest outflowing gas in IC~5063, the high line ratios  are explained by the gas being clumpy and having kinetic temperatures in the range 20 -- 100 K and densities between $10^5$ and $10^6$ cm$^{-2}$ \citep{Oosterloo17}. These conditions may be indicative for  \target.   Similar situations have now been found in a number of objects across a range of radio jet powers, such as  PKS~1549$-$79 \citep{Oosterloo19}, NGC~3100 \citep{Ruffa22}, the Teacup \citep{Audibert23}, and B2~0258+35 \citep{Murthy25}.   

For the regions with high CO ratios that are well outside the visible radio structures, it is more difficult to find an unambiguous explanation for the observed physical conditions of the gas. 
In two of the objects mentioned above (B2~0258+35 and the TeaCup), high line ratios in the region perpendicular to the direction of the radio jets have  been observed and were explained as the signatures of the effect of the latterly expanding cocoon of shocked gas created by the jet-ISM interaction.   Although  radio jets with similar power as  those in \target\  ($P_{\rm jet} \ge 10^{45}$ erg s$^{-1}$) can  directly punch  through the circumnuclear ISM more rapidly than low-power jets,  simulations showed that in powerful sources, a cocoon of shocked gas is also created that expands in the direction perpendicular to the jet \citep{Mukherjee16}. Therefore, we conclude that at least part of the gas with a high line ratio that is observed in the direction orthogonal to the radio axis can be created by such a cocoon.  An alternative explanation could be that the X-ray emission that is observed in this region excites the CO gas to higher excitation levels, but this would require high X-ray fluxes and high densities \citep{Lamperti20}. The radial size of the affected regions in the two low-power sources B2~0258+35 and the TeaCup is about 0.7 kpc, which is smaller than the $\sim$2 -- 5 kpc observed in \target. Interestingly, new ALMA observations of powerful AGN (Morganti et al.\ in prep.) show regions  with high line ratios that extend over similar spatial scales (radius of a few kpc) to those observed in \target. 

On the other hand, \target\ is embedded in an on-going galaxy interaction or merger, as is evidenced by the unsettled overall nature of the CO on large scales. Therefore, in addition to the impact of the AGN, the  CO excitation might be affected by turbulence and dissipation as the accreted gas settles into a more stable configuration. For example,
this mechanism might explain
the region of enhanced excitation beyond the N lobe in the direction of the radio axis, which is otherwise difficult to reconcile with the effects of jet-cloud interactions \citep[e.g.][]{Mukherjee16}. In this case, we would expect enhanced CO excitation to be a common feature of the central regions of gas-rich mergers. Therefore, future spatially resolved CO line ratio studies of non-AGN mergers are required to confirm or refute this idea.

\subsection{The energetics of the near-nuclear outflow}
\label{sec:Energetics}

As mentioned above, in the  sub-kpc central region, the broad CO line suggests an outflow. This is also supported by the high line ratios that are indicative of high excitation  and/or optically thin conditions.  It is difficult to derive the properties of the outflow because it is only detected in \coTwo\ and \coThree, not in \coOne. The non-detection of \coOne\ is one of the surprising results because of the extreme conditions this implies.    Using the upper limit to the CO(1-0) flux integral for the centre derived in Sect.\ \ref{sec:lineRatios}, we derived an upper limit to the mass of the molecular gas in the outflow (in the central kpc) of $M_{\rm out} < 1.9 \times 10^8$ \msun, assuming $\alpha_{\rm CO} = 0.89$ \msun (K \kms\ pc$^2$)$^{-1}$. 

The mass outflow rate can be estimated using $\dot{M} \sim M_{\rm out} \times v_{\rm out} /r_{\rm out}$ with $v_{\rm out}$ the outflow velocity and $r_{\rm out}$ the extent of the outflow.
We took $v_{\rm out}$ to be in the range between 200 and 300 \kms, the former representing an estimate of the bulk of the outflow  and the latter the maximum velocity with respect to the systemic velocity at which we observe gas. A  conservative estimate for  the radial extent $r_{\rm out}$  is about 1 kpc, that is, about one beam. With these assumptions, we derived an upper limit on the  mass outflow rate (by taking $v_{\rm out} = 300$ \kms) of  $\dot{M} \lesssim   $ 20 \msunyr. 
As stated above, we suspect that the true mass of the molecular outflow is not much lower than our upper limit, and hence, that the actual mass outflow rate is only slightly lower than 20 \msunyr.
If this is indeed the case, the mass outflow rate for the molecular gas is much higher than the $\dot{M} \sim 1$ \msunyr\  measured for the warm ionised gas component (\citealt{Santoro20}). 
Using our estimate of the mass outflow rate of the cold molecular gas, we derived the kinetic power, estimated as  $\dot{E}_{\rm kin} = \frac{1}{2}\dot{M}(v^2_{\rm out}+3\sigma^2)$, and find  
$\dot{E}_{\rm kin} \lesssim 1 \times 10^{42}$ \ergs\ assuming  $\sigma = 85$ \kms. 

The  jet power (a few $\times 10^{46}$ \ergs\ derived in \citealt{Morganti21}) is much higher   than this kinetic power, and   therefore, there is enough energy on kpc scales in the radio plasma to accelerate, heat, and disperse the gas and to produce the cocoon of shocked gas. 
As mentioned above, however, although the impact of the radio plasma is clear at large radii, on the small scales at which the outflow is observed, we cannot exclude that it is driven by the impact of radiation from the optical AGN. The high bolometric luminosity ($L_{\rm bol} \sim 2.5 - 4 \times 10^{45}$ \ergs, \citealt{Santoro20}) implies that radiation pressure might be a possible driver of the outflow. The uncertainties connected to many parameters (e.g.\ the outflow launching radius, the mass of the galaxy inside that radius, and the dust properties) mean, however,  that we cannot precisely quantify the importance of the radiation pressure.  

In summary, the results above suggest that while the jet evolves and grows, the impact of the radio plasma changes: In the first phase, the jet/lobe directly interacts with dense clouds in the near-nuclear regions on sub-kpc scales, and it might accelerate them to several 100 km s$^{-1}$. Then, when the jet expands to larger scales, it interacts with lower-density gas clouds that cannot cool quickly enough to avoid shredding and destruction by hydrodynamic instabilities associated
with the hot wind behind the jet-induced shock. On  larger scales, however, the interaction with the radio source is still visible in the enhanced CO excitation and in the velocity dispersion that results from a milder and less destructive interaction with the more slowly expanding jet cocoon.  Thus, part of the energy that is deposited by the interaction can go into affecting the conditions of the gas in two ways: maintaining the turbulence, as indicated by the high-velocity dispersion, and increasing the excitation temperature of the gas.

 \begin{figure}
   \centering
      \includegraphics[angle=0,width=9cm]{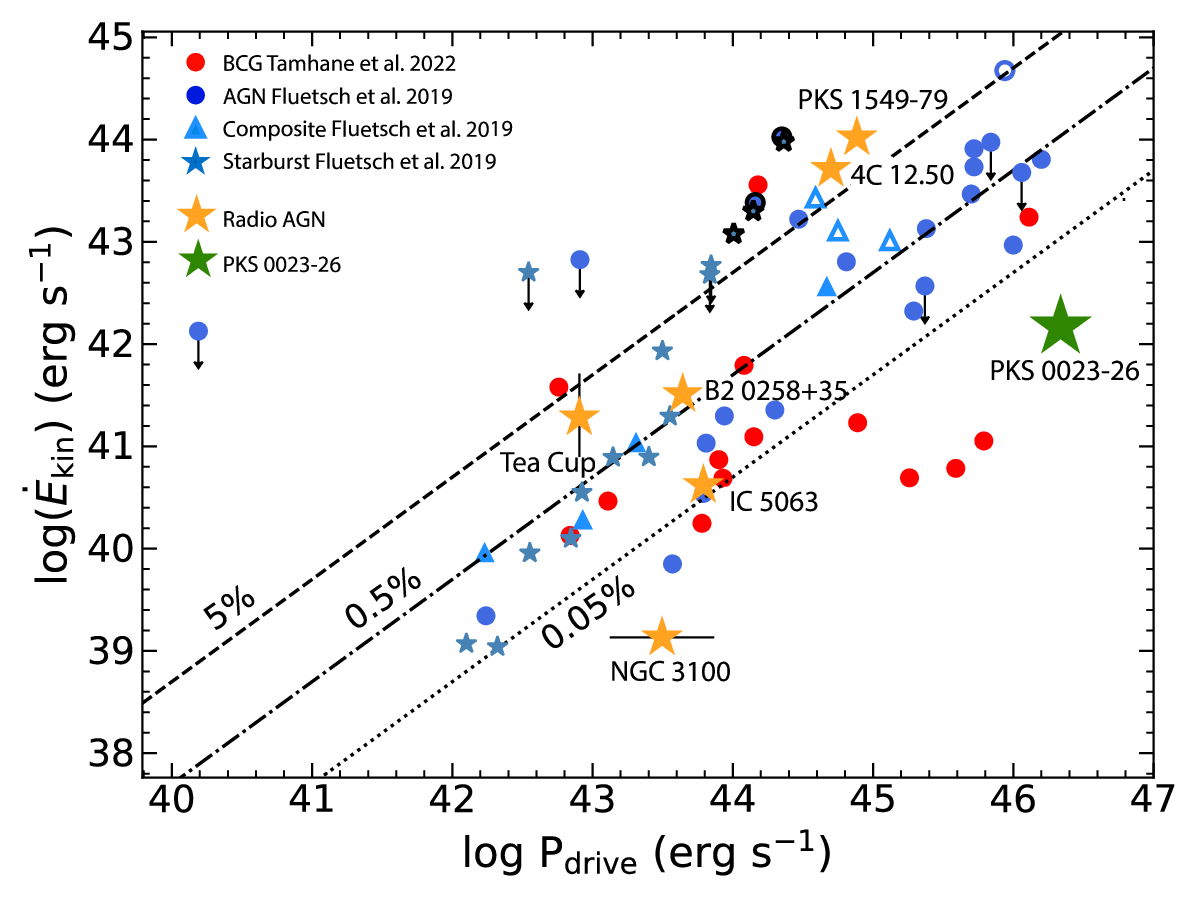}
   \caption{Modified from  Fig.\ 9 of \cite{Tamhane22}, we show the relation between the kinetic power of flows (from cold molecular gas) and the mechanical power for BCGs together with AGN (composite and starburst), taken from \citet{Fluetsch19}. For the AGN, the x-axis represents the bolometric AGN luminosity. \target\ and a few radio AGN (see Sect. \ref{sec:ComparisonAGN}) where outflows of cold molecular gas are detected were added to the plot and are indicated with large stars (see Sect. \ref{sec:interaction} for details on \target).
   The black lines represent flow powers of 5\% (dashed), 0.5\% (dash–dotted), and 0.05\% (dotted) that of the bolometric AGN luminosity. 
}
\label{fig:Energetics}
\end{figure}
\noindent

\subsection{Impact and comparison with other AGN}
\label{sec:ComparisonAGN}

The energetics derived for \target\ suggest that the kinetic energy that is transferred to the ISM is only a small fraction of the energy that is available in the radio jets. This would suggest that although we see a clear coupling between the radio plasma and the cold molecular gas, most of the energy of the jets escapes from the galaxy, possibly to the CGM.
This has been reported for other AGN in the literature, and we illustrate this in Fig. \ref{fig:Energetics} , modified from the figure presented by  \cite{Tamhane22}. These authors compared the energetics derived from the cold molecular gas observed in BCG located in cool-core clusters with those of galaxies hosting AGN with wind- or radiation-driven outflows (also in cold molecular gas) from \cite{Fluetsch19}. Depending on the type of AGN,  they used the mechanical power of the jet (for the radio loud AGN) or the bolometric luminosity for the optical AGN for $P_{\rm drive}$ . 

In this plot, we add next to \target galaxies for which jet-driven cold molecular outflows were observed and their energetics derived. In particular, we included two high radio power AGN (4C~12.50 and PKS~1549--79; \citealt{Holden24,Oosterloo19}, respectively) and  four lower-luminosity radio AGN (IC~5053 \citealt{Oosterloo17,Holden23};  B2~0258+35 \citealt{Murthy22,Murthy25}, the TeaCup \citealt{Audibert23}; and NGC 3100 \citealt{,Ruffa22}) that all represent young radio galaxies. For these radio AGN, including \target, we used the jet kinetic power as $P_{\rm drive}$. The general conclusion does not change when the bolometric luminosity is used for the objects that also host an optical AGN (as is the case of \target).

The figure shows that the kinetic energy fluxes of molecular gas in most systems lie near or below a few percent of the AGN or the starburst power, regardless of the driving mechanism of the outflows,    as was also reported by \cite{Audibert25}. Interestingly, \target\ appears to be well below the 0.05\% line, like some BCG, and lower than other similarly powerful radio galaxies.
Considering the more advanced evolutionary stage of \target, we suggest that the coupling between the energy and the ISM may have been high for a short initial period when the radio jet interacted strongly with the higher-density gas in the central regions of the galaxy before it escaped into the extended halo of the galaxy. If this is the case, it would agree with  predictions from  simulations \citep{Mukherjee18a}.

From the results described above, we conclude that \target\ shows the effect of the two different types of feedback: ejective in the central sub-kpc region and possibly at the location of the (northern) radio lobe, and preventative on the scale of a few kpc, in a way  similar to what is seen in clusters, where the molecular gas appears to be wrapped around the radio lobes. 

It is interesting to note that gas cocoons created by the interaction of the radio plasma with the ISM are seen in a growing number of objects. For example, \cite{Girdhar24} found similar features in radio-quiet type 2 quasars where extended radio emission, albeit much weaker than in radio galaxies, has been observed. This  confirms  that low-power jets can also affect their surrounding medium in a ways that can go beyond producing fast gas outflows.

\section{Conclusions} 
\label{sec:conclusions}

We have used multiple CO transitions (\coOne, \coTwo\, and \coThree) to investigate the kinematics and physical conditions of the cold molecular gas around a powerful, young radio galaxy, \target.
The results show that the information from the CO line ratios can complement and expand the view of the AGN impact on the cold gas as derived solely from the gas kinematics.  
Although the results from the three transitions show similarities to first order, clear differences can be noted in their relative intensities at different locations. This suggests different conditions of the gas. This is particularly extreme in the case of \coOne, which is undetected in the central regions that are coincident with the radio core. This is the same location as that at which \coThree\ shows its peak intensity. 

The region with the highest $R_{32}$ line ratios was found to be aligned with the radio emission north-west of the nucleus, where  relatively high velocity dispersions were also  measured. This clearly demonstrates the impact of the radio source on the excitation and kinematics of the molecular gas, and it might result from interaction of the gas with the slowly expanding  cocoon that is created by the jet as it finds its way through the ISM. More destructive interactions with the radio source likely occur at the site of the N radio lobe itself, however, which appears to have very little CO emission. At this location, the CO emission is not only very faint (see Fig. \ref{fig:channels}), but also broad in velocity. 

\target\ is one of a growing number of AGN that show regions of extreme CO line ratios that also extend in a direction  perpendicular to the radio jet. To first order, this is consistent with the predictions of numerical simulations, where the jet-ISM interaction can produce a cocoon of shocked gas that can extend in the direction orthogonal to the jet. This affects the physical condition of the gas well outside the visible radio source.  A region of high CO excitation
($R_{\rm 32} > 0.6$) that extends up to $\sim$2 kpc beyond the N radio lobe in the jet direction is difficult to reconcile
with this explanation, however. 
In this context, it is notable that \target\ is relatively gas rich for an early-type galaxy and shows clear signs of interactions with neighbouring galaxies. This suggests an alternative explanation for the high line ratios in terms of the turbulence and dissipation induced as the accreted gas settles following a galaxy merger or interaction. 

From the kinematics, excitation, and the distribution of the cold molecular gas, we were able to trace different types of interaction between the (radio) AGN and the gas. The fastest gas outflows are confined to the central kpc, and the effects at larger radii are milder. As suggested by \citet{Morganti21}, the AGN  feedback may appear in different forms during the evolution of the radio jet. 
Most importantly, we confirmed that the impact of the AGN is not necessarily only seen in the kinematics, but rather in the line ratios, and their spatial distribution are also powerful diagnostics. This introduces additional complexity in our attempt to describe and quantify AGN feedback.  

Another important finding is that although the effects of the interaction   between the radio plasma and the ISM are clear, the energy that is transferred from the AGN to the gas is only a small fraction of the energy available in the jet. This adds to a growing body of evidence that now expands to high redshift  through the results obtained with JWST (e.g.\ \citealt{Cresci23,Roy24,Eugenio25}), and it implies that much of the jet energy is likely to escape the galaxy to affect  its environment on larger scales.

\begin{acknowledgements}
We thank Tim Davis for his input on the CO line ratios of early-type galaxies and we thank an anonymous referee for very useful input. 
This paper makes use of the following ALMA data: ADS/JAO.ALMA\#2022.1.01498.S. ALMA is a partnership of ESO (representing its member states), NSF (USA) and NINS (Japan), together with NRC (Canada), MOST and ASIAA (Taiwan), and KASI (Republic of Korea), in cooperation with the Republic of Chile. The Joint ALMA Observatory is operated by ESO, AUI/NRAO and NAOJ. 
\end{acknowledgements}

\newpage
\appendix 
\begin{appendix}
\section{Absorption against the N radio lobe}
\label{sec:Abs-Appendix}	

\begin{figure}[h]
   \centering
      \includegraphics[angle=0,width=8cm]{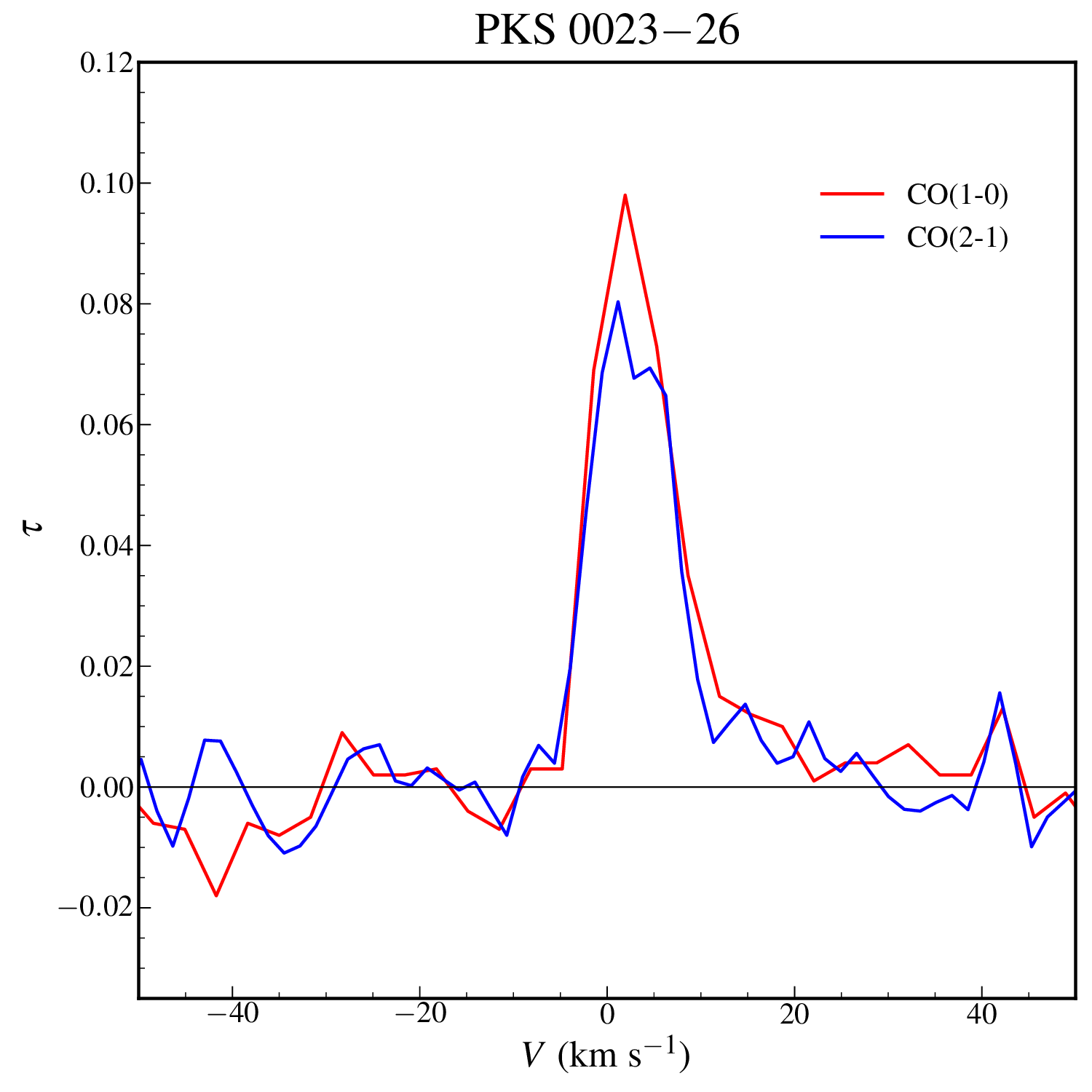}
   \caption{Profile of the absorption in two transitions (\coTwo\ and \coOne) in units of $\tau$ at the location of the peak continuum of the N lobe obtained from the highest spatial  (i.e., where most of the emission is resolved out) and spectral resolution cube. 
  }
              \label{fig:absorption}
    \end{figure}

As was the case in the \coTwo\ observations of \cite{Morganti21}, absorption is  present in the \coOne\ data at the location of the peak of the continuum emission of the northern lobe. No absorption is detected in the \coThree\ data, but this is at least in part due to the faint continuum flux (26 \mJybeam) of the peak of the northern lobe in the \coThree\ data compared to that at the other frequencies (see Appendix \ref{sec:AppendixCont}).  We characterise the absorption using the highest resolution cubes described in Sect. \ref{sec:observations} and Table \ref{tab:obs}.
Figure \ref{fig:absorption} shows the two absorption profiles  plotted in  optical depth $\tau$.  The profiles are extracted from the high spatial- and velocity resolution \coOne\ and \coTwo\ cubes, given the small velocity width of the absorption.

As in the case of \coTwo, the \coOne\ absorption is narrow with a FWHM $\sim$10 \kms,  roughly centred on the systemic velocity. This is consistent with what was reported by \cite{Morganti21}, although those values were derived from a lower velocity resolution cube.  The depth of the \coOne\ absorption is almost 12 \mJybeam\ against the continuum peak flux of 123 \mJybeam. As discussed in \cite{Morganti21}, the narrow width of the absorption suggests that it is likely produced by a single giant molecular cloud \citep{Miura21} observed against the small, compact component in the N lobe which has been detected  in VLBI and that has a size of about 100 pc \cite{Tzioumis02}. The results from  Gaussian fits to the \coOne\ and \coTwo\ profiles are given in Table \ref{tab:absFit}. 

From the detection of absorption in two transition, in principle, one can estimate the excitation temperature \Tex\ using
\[
\frac{\int \tau_{21} dv}{\int \tau_{10}dv} = 2 \frac{1-e^{-h\nu_{21}/kT_{\rm ex}} }{e^{h\nu_{10}/kT_{\rm ex}} - 1} .
\]
Using the observed values for the \coOne\ and \coTwo\ absorption, we find $T_{\rm ex} = 5 \pm 1$ K. However, as described by \cite{Rose24}, the fact that the background continuum may have different sizes at different frequencies, as well as  the complex hierarchical structure of the molecular clouds, make the interpretation of optical depths ratios very uncertain.
Using \Tex\ = 5 K and a CO abundance relative to H$_2$ of $10^{-4}$, we find that the H$_2$ column density is $N_{{\rm H}_2} = 2.0 \pm 0.5 \cdot 10^{19}\ {\rm cm}^{-2}$. This is in the range of column densities observed by \cite{Rose24} in a number of radio galaxies.

\begin{table}[h]
	\caption{Parameters of Gaussian fits to the absorption profiles}
	\label{tab:absFit}
	\begin{center}
	\begin{tabular}{ l c c}
		\hline \hline
        Parameter & \coOne  & \coTwo  \\
        \hline
        Peak $\tau$       &  0.099 $\pm$ 0.007 & 0.083 $\pm$ 0.005 \\
        FWHM \ \  (\kms)            &  9.4 $\pm$ 0.8  &  10.4 $\pm$  0.7\phantom{9}\\
        $V_{\rm cen}$ \ \ (\kms)    &  2.6 $\pm$ 0.3  &  2.8  $\pm$  0.3  \\
        $\int \tau\,dv$ \ \ (\kms)  &  0.99 $\pm$ 0.20 & 0.92 $\pm$  0.15 \\
\hline

		\end{tabular}
      \end{center}
\end{table}

\FloatBarrier

\section{Continuum properties}
\label{sec:AppendixCont} 

\begin{figure}
   \centering
      \includegraphics[angle=0,width=8cm]{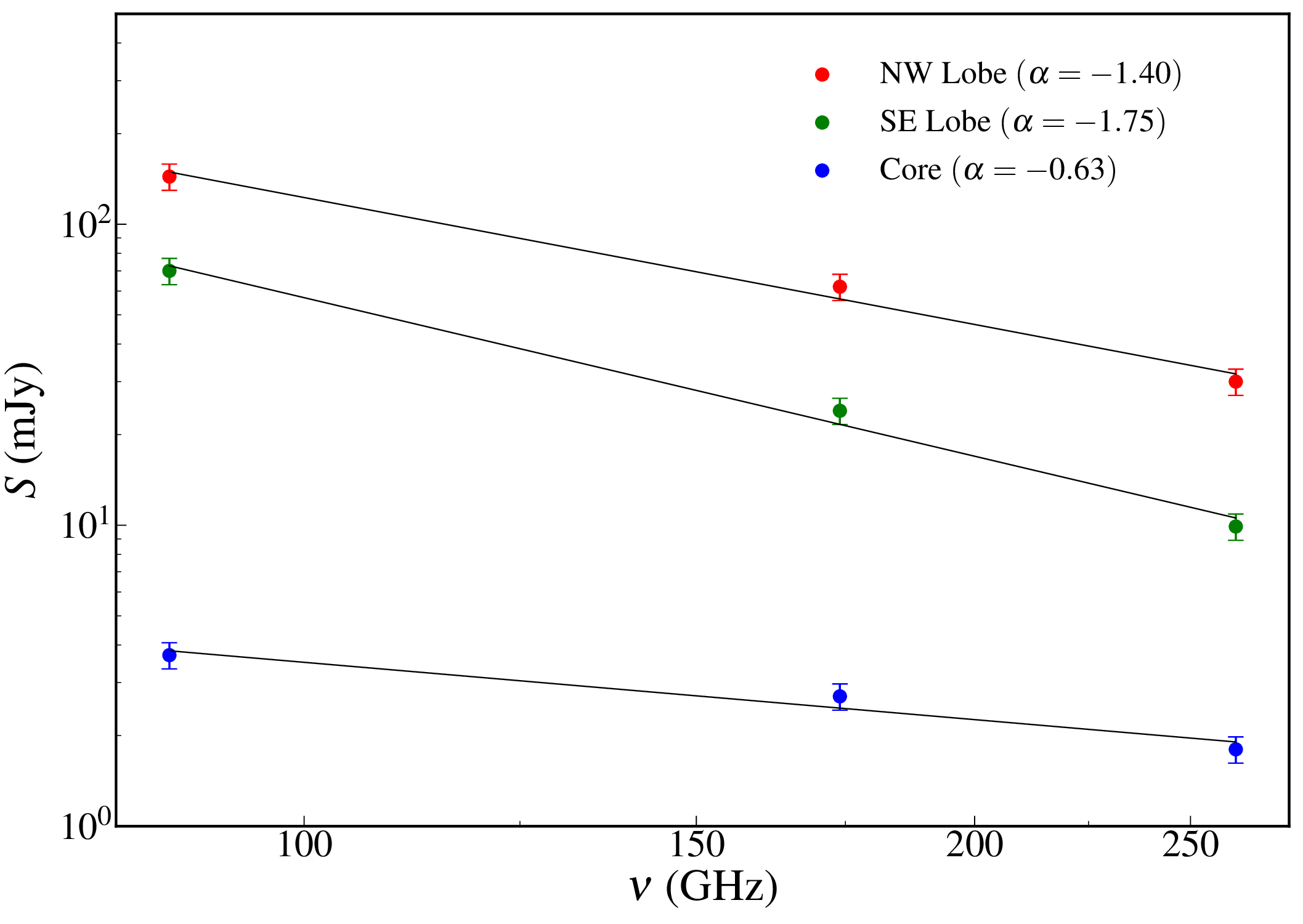}
   \caption{Radio spectra of the three continuum components. The black lines indicate the fitted power law spectra.
  }
              \label{fig:specindex}
    \end{figure}

The  continuum images made at the three frequency bands (see Table \ref{tab:obs}) all show the same structure of a core and two lobes (see Fig.\ \ref{fig:molecularGas} and \cite{Morganti21}). Table \ref{tab:fluxes} and Fig.\ \ref{fig:specindex} give the integrated flux for the three components and their spectral indices $\alpha$ (defined through $S\propto \nu^\alpha$) as derived from these fluxes. The spectral indices of the lobes are quite steep and steeper than seen at lower frequencies \citep{Morganti21}. This trend, which suggests the presence of a spectral break at relatively high frequencies, is  often  seen in Compact Steep Spectrum radio galaxies similar to \target, as shown in \cite{Murgia99}. According to \cite{Murgia99}, this confirms the idea that these radio galaxies are genuinely young.

 \begin{table}
\caption{
Continuum flux densities and spectral indices of the three components derived from the ALMA observations. 
}

\begin{center}
\begin{tabular}{lcccccc} 
\hline\hline 
   &   $S_{\rm 87\ GHz}$ & $S_{\rm 174\ GHz}$ & $S_{\rm 262\ GHz}$ &$\alpha$\\
      &  (mJy) &  (mJy) &  (mJy) \\
\hline
N lobe     &  144   & 62   & 30  & $-1.40 \pm 0.14$ \\
SE - lobe  &   70   & 24   & 9.9 & $-1.75 \pm 0.16$ \\
Core       &    3.7 & 2.7  & 1.8 & $-0.63 \pm 0.15$\\

\hline
\end{tabular}

\end{center}
\label{tab:fluxes}
\tablefoot{The error in the fluxes is assumed to be 10\%.}

\end{table}

\end{appendix}

\end{document}